\documentclass[preprint,11pt,authoryear,nonatbib]{elsarticle}
\usepackage{booktabs}
\usepackage{colortbl}
\usepackage{xcolor}
\usepackage{graphicx}
\usepackage[T1]{fontenc}
\usepackage[utf8]{inputenc}

\usepackage[T1]{fontenc}
\usepackage{newunicodechar}

\newunicodechar{́}{\'} 
\newunicodechar{̌}{\v} 

\usepackage{subcaption}
\usepackage{comment}

\newcommand{\cellcolorByValue}[1]{%
    \ifdim#1 pt<0pt\relax
        \cellcolor{red!50}%
    \else
        \ifdim#1 pt=0pt\relax
            \cellcolor{blue!25}%
        \else
            \cellcolor{green!50}%
        \fi
    \fi
    #1%
}

\makeatletter 
\let\c@author\relax
\makeatother

\usepackage{textgreek}
\usepackage[margin=2.5cm]{geometry}
\graphicspath{{figure/}}
\usepackage{color,soul}
\usepackage{amsmath}

\PassOptionsToPackage{style=numeric-comp,sorting=none}{biblatex}
\usepackage[style=numeric-comp, sorting=none]{biblatex} 
\title{Coupling Anisotropic Curvature and Nematic Order: Mechanisms of Membrane Shape Remodeling}

\author[add1]{Yoav Ravid }
\author[add2]{Samo Peni\v{c}}
\author[add2]{Luka Mesarec}
\author[add1]{Nir S. Gov}
\author[add3]{Veronika Kralj-Igli\v{c}}
\author[add2]{Aleš Iglič}
\author[add2]{Mitja Drab \corref{cor1}}
\cortext[cor1]{Corresponding author: mitja.drab@fe.uni-lj.si}

\address[add1]{Department of Chemical and Biological Physics, Weizmann Institute of Science, Rehovot 7610001, Israel}
\address[add2]{Laboratory of Physics, Faculty of Electrical Engineering, University of Ljubljana, Ljubljana, Slovenia}
\address[add3]{Laboratory of Clinical Biophysics, Faculty of Health Sciences, University of Ljubljana, 1000 Ljubljana, Slovenia}

\begin{document}

\begin{abstract}
This study theoretically investigates how anisotropic curved membrane components (CMCs) control vesicle morphology through curvature sensing, nematic alignment, topological defects and volume constraints. By comparing arc- and saddle-shaped CMCs, we identify a rich spectrum of steady-state phases. For fully CMC-covered vesicles, arc-shaped components drive a pearling-to-cylinder transition as nematic interactions strengthen, while on partially CMC-covered vesicles the saddle-shaped CMCs stabilize necks between the convex regions of bare membrane. We map the steady-state shapes of vesicles partially covered by arc- and saddle-shaped CMCs, exposing how different vesicle shapes depend on the interplay between nematic interactions and volume constraints, revealing several novel phases. By investigating the in-plane nematic field, we find that topological defects consistently localize to high-curvature regions, revealing how intrinsic and deviatoric curvature effects cooperate in membrane remodeling. These findings establish a unified framework for understanding how proteins and lipid domains with anisotropic intrinsic curvature shape cellular structures---from organelle morphogenesis to global cell shape.
\end{abstract}



\maketitle


\section{Introduction}\label{sec:intro}

Membranes are fundamental structures in cells that act as barriers between compartments and facilitate essential biological processes. These membranes dynamically change shape to accommodate various cellular functions, influenced by interactions with membrane proteins and lipids \cite{simunovic2015physics,has2021recent,johnson2024protein,has2022insights,gomez2016anisotropic}. Understanding how these shape changes occur is crucial to understanding the intricate workings of a cell. An integral part of lipid membranes is curved membrane components (CMCs), aggregates of proteins or lipid rafts \cite{schamberger2023curvature} that can move laterally along the membrane surface, bend the membrane locally, and are sensitive to local membrane curvature.

There is ample experimental evidence to suggest that CMCs have different curvature preferences along different directions, which means they are intrinsically anisotropic. A notable example of anisotropic CMCs is the family of BAR domain proteins \cite{simunovic2015physics,gallop2005bar}. Amphiphysin 1, an N-BAR protein, concentrates in membrane nanotubes and induces tubulation \cite{simunovic2019curving, simunovic2016physical}. Another example is IRSp53, a protein that contains an I-BAR domain and exhibits a strong preference for negatively curved membranes and is found on the inner leaflet of membrane nanotubes pulled from giant unilamellar vesicles \cite{johnson2024protein,tsai2022activated,drab2019inception,prevost2015irsp53}. Coupling between the non-homogeneous lateral distribution of membrane components and the local anisotropic membrane curvature has also been recently indicated in the Golgi, where some of the membrane components are concentrated on the bulbous rims of the Golgi vesicles and where the difference between the two principal membrane curvatures is very large \cite{iglicFEBS04}. Similar phenomena have also been observed in photoreceptor discs \cite{corbeil2001rat,chandler2008intrinsic}, endoplasmic reticulum shapes \cite{guven2014terasaki}, and flattened endovesicles of the erythrocyte membrane \cite{hagerstrand2004endovesicle}. These examples indicate that the coupling between the non-homogeneous lateral distribution of generally anisotropic membrane components may be a general mechanism stabilizing highly curved membrane structures and demonstrate the importance of anisotropic membrane components in driving cellular processes \cite{kralj2000stable, simunovic2016physical,li2019nanostructure,gomez2016anisotropic,zimmerberg2006proteins,kabaso2012role,campelo2010modeling,idema2019interactions,kralj1999free, schamberger2023curvature,kralj-iglic2020minimizing,sadhu2021modelling,sadhu2023minimal}. Sometimes, several types of curved-like proteins may be working in a coordinated manner to induce membrane morphologies, and tubulation can be studied in the context of a mixture of positive and negative curvature proteins \cite{kumar2022membrane}.

Numerical simulations of membranes that are not bound by constraints of axisymmetrical shapes have been a valuable tool in understanding the coupling between membrane curvature and curved membrane components (CMCs) that can drive feedback loops by curvature sensing and clustering. In a comprehensive review by Ramakrishnan et al., shape transformations of membranes with an in-plane nematic field were explored for partial and full coverage fraction \cite{ramakrishnan2014mesoscale}. Similarly, an open source tool was developed to study the effects of nematic interactions used for the analysis of real membranous systems \cite{pezeshkian2024mesoscale}. 
 
However, studies of membrane deformation that include anisotropic CMCs in combination with the constraint of fixed volume remain poorly understood and to our knowledge, lacking. Here, we expand upon our previous work \cite{ravid2024numerical} and present new numerical results obtained from Monte Carlo simulations of triangulated, closed membrane shapes using anisotropic CMCs which interact nematically. We explore how the spontaneous curvature of CMCs (arc-shaped or saddle-shaped), CMC concentration, and strength of nematic ordering, together or in the absence of constraints on volume, influence the overall membrane shape. We first start with vesicles fully covered with arc- and saddle-like CMCs and find that the steady-state shapes are mostly cylindrical and globally saddle-shaped, respectively. We find a pearling-to-cylinder transition that occurs at low nematic interaction strength. When the vesicles are only half covered by non-interacting arc-shaped CMCs, with constraints of a constant volume, we find that there is an accelerated tendency toward global prolate shapes in comparison with empty vesicles, which can sustain an oblate phase up to larger relative volumes. When nematic interaction is present, we explore the phase diagram in the space of relative volume and CMC interaction strength, and characterize a boomerang-like phase that marks the transition from the oblate to the prolate phase. For saddle-like CMCs, we find that full coverage and volume constraints can lead to global saddle shapes with the formation of singular membrane protrusions. When the concentration of CMCs is decreased, the volume-binding strength phase diagram features stomatocyte, oblate-like, and protrusive phases. We find that at small saddle-like CMC concentrations, these have a tendency to aggregate in the necks between two convex parts of the membrane, which leads to the overall elongation of the vesicles. We find that simple physical interactions between anisotropic CMCs and their associated in-plane nematic fields can drive diverse membrane transformations, such as tubule formation and pore creation \cite{fovsnaric2005influence}.

\section{The model}
\subsection{Membrane-protein interactions}

\begin{figure}
    \centering
    \includegraphics[width=\textwidth]{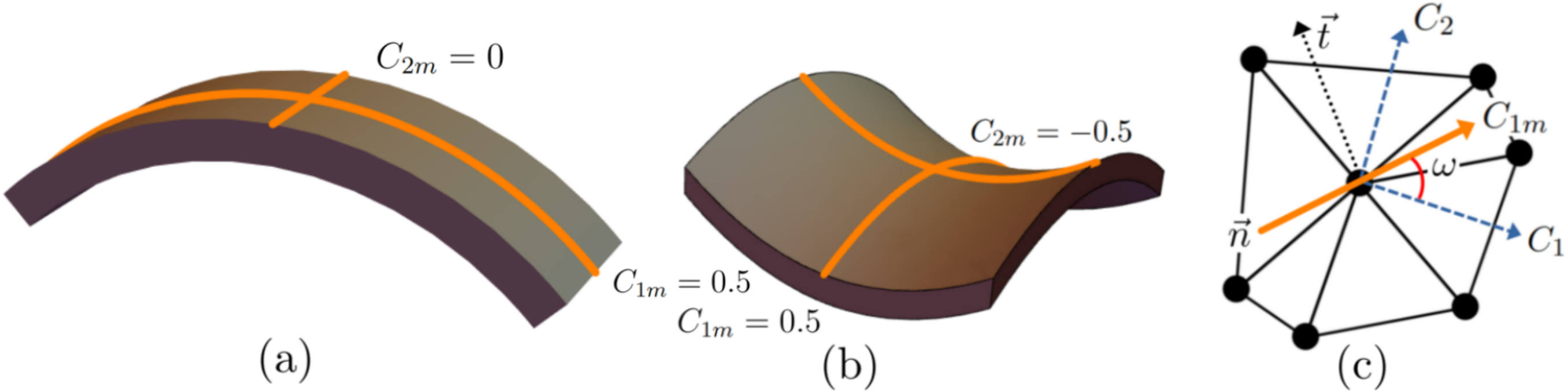}
    \caption{(a) Example of an arc-shaped CMC with $H_m=D_m=0.25$ ($C_{1m}=0.5$ and $C_{2m}=0$). (b) Example of a saddle-shaped CMC, with $H_m=0$, $D_m=0.5$. (c) The mismatch between the principle curvatures of a membrane protein and the local curvature of the triangulated membrane. Here, $\vec{n}$ is the orientation of the CMC and $\vec{t}$ is its perpendicular direction in the local tangent plane.}
    \label{fig:fig1}
\end{figure}

The theoretical framework is grounded in the derivation of the membrane bending energy and anisotropic inclusion interactions \cite{iglivc2007,iglivc2005role,kralj2002deviatoric,kralj2012stability}. The model describes the membrane as a 2D anisotropic continuum surface. The elastic energy of a small membrane section is lowest when its principal curvatures $C_1$ and $C_2$ align with the intrinsic curvatures of the anisotropic CMCs, $C_{1m}$ and $C_{2m}$, at that location. The level of alignment can be quantified by comparing the orientations of the local membrane curvature tensor $C$ and the intrinsic membrane curvature tensor $C_m$. 
$$
C = \begin{pmatrix} C_1 & 0 \\ 0 & C_2 \end{pmatrix}, \quad C_m = \begin{pmatrix} C_{1m} & 0 \\ 0 & C_{2m} \end{pmatrix}.
$$
These two tensors---describing the directions of the membrane and CMC principal curvatures, respectively---are generally rotated relative to each other by an angle $\omega$ in the tangent plane. The mismatch tensor is defined as $M = R C_m R^{-1} - C$, where $R$ is the rotation matrix that quantifies the angular disparity
\cite{kralj2002deviatoric,iglivc2005role,kralj2002deviatoric,kralj2012stability}:
$$
R = \begin{pmatrix} \cos \omega & -\sin \omega \\  \sin \omega & \cos \omega \end{pmatrix}.
$$
A CMC reorients in the tangent plane by $\omega$ radians to match the membrane curvature and minimize energy, reflecting the energetic cost of deformation (see Figure \ref{fig:fig1}(b)). The approximate elastic energy, $E_1$, is expressed as a series expansion in the independent invariants of the mismatch tensor $M$. Using the trace and determinant of $M$ as invariants yields the following expression \cite{iglivc2005role,kralj2002deviatoric,kralj2012stability,kralj1996shapes}:
\begin{equation}\label{eq:eq1}
    E_1 = \int \frac{K_1}{2} (\mathrm{Tr} M)^2 + K_2 \mathrm{Det} M \,dA,
\end{equation}
integrated across the entire vesicle. Substituting the bending moduli $K_1$ and $K_2$ gives \cite{iglivc2005role,kralj2002deviatoric,kralj2012stability,kralj1999free}:
\begin{equation}\label{eq:eq2}
E_1 = \int (2K_1 + K_2)(H - H_m)^2 - K_2 \left( D^2 - 2DD_m \cos 2\omega + D_m^2 \right)\,dA,
\end{equation}
where $D = (C_1 - C_2)/2$ is the curvature deviator, while $H_m = (C_{1m} + C_{2m})/2$ is the intrinsic mean curvature, and $D_m = (C_{1m} - C_{2m})/2$ the intrinsic curvature deviator. The curvature deviator $D$ can be nondimensionalized as $d = DR$, where $R$ is the radius of a sphere with the same volume $V$ and area $A$ as the vesicle. In the case of an isotropic membrane where $D_m=0$, it becomes evident that Eq. (\ref{eq:eq2}) is equal to the Helfrich bending energy density \cite{helfrich1973} described by $E_b = k^2_c (2H - C_0)^2 + k_G K$, where $H = (C_1 + C_2)/2$ is the mean curvature, $C_0$ is the spontaneous curvature, $K = C_1C_2$ is the Gaussian curvature, and $K_1=k_c$ and $K_2 = k_G$, where $k_c$ and $k_G$ are the bending and splay moduli, respectively \cite{iglivc2005role,kralj1996shapes,kralj2012stability}. To compute the principal curvatures at each vertex of a triangulated membrane mesh, we adapt the method of Ramakrishnan et al. \cite{ramakrishnan2014mesoscale} with several key modifications (see Appendix \ref{app:code details}). While our method adopts the same CMC–CMC interaction model as Kumar et al. \cite{kumar2022review}, it differs fundamentally in the formulation of membrane bending energy (Eq. \ref{eq:eq1}). 

\subsection{Isotropic protein-protein binding}

Two neighboring CMCs may bond to each other on the membrane, thereby lowering the overall energy:
\begin{equation}\label{eq:eq3}
    E_2 = \begin{cases} - w & i,j\text{ contain CMC} \\ 0 & \text{else} \end{cases}
\end{equation}
Here, the type and binding strength $w$ are vertex properties that reflect local aggregation. Throughout this paper, we refer to this interaction as isotropic bonding. Such bonding is commonly observed in surfactants and proteins, as it results from spontaneous self-assembly on the membrane driven by hydrophobic and other intermolecular effects \cite{deleu2003interaction,fosnaric2019theoretical}.

\subsection{Nematic protein-protein interactions}

Anisotropic CMCs exhibit nematic interactions similar to those observed in liquid crystals. The energy between neighboring CMCs will be lower if their principal curvatures are aligned, which we model using Frank's free energy density for nematic liquid crystals \cite{frank1958liquid}:
\begin{equation}\label{eq:eq4}
\begin{aligned}
E_2 &=  \int \left(\frac{k_G'}{2}\left(\nabla\cdot\vec{n}\right)^2 + \frac{k_c'}{2}\left(\nabla\cdot\vec{t}\right)^2\right) r_v \,dA
\end{aligned}
\end{equation}
Here, $\nabla$ represents the covariant derivative on the curved surface, $\vec{n}$ denotes the orientation of the inclusion, and $\vec{t}$ signifies its perpendicular direction in the local tangent plane. The constants $k_G'$ and $k_c'$ correspond to the elastic constants governing in-plane nematic interactions. The variable $r_v$ takes a value of 1 if a CMC is present on a vertex; otherwise, it is 0.

A discrete form of this energy is employed to facilitate implementation in the Monte-Carlo (MC) simulations. If we assume a one-constant approximation ($k_G'=k_c'$), Eq. \ref{eq:eq4} can be rewritten in a way that makes the implementation suitable for MC simulations (Lebwohl-Lasher model) \cite{lasher1972monte}:
\begin{equation}\label{eq:eq5}
    E_2 \approx - \int \sum_{k=1}^{N} \epsilon_{LL}^{k,k} \sum_{i>j}\left(\dfrac{3}{2}(\vec{n_i} \cdot \vec{n_j})^2 -\dfrac{1}{2} \right)r_v \,dA. 
\end{equation}
Here, $\epsilon_{LL}$ is the strength of the nematic interaction. The sum $\sum_{i>j}$ is over all the nearest neighbor (i, j) vertices on the triangulated grid, promoting alignment among the neighbouring orientation vectors. An even simpler form of this approximation for the in-plane orientational field is the XY model on a surface \cite{ramakrishnan2010monte}:
\begin{equation}\label{eq:eq6}
E_2 \approx -w \int \sum_{i>j}(\vec{n}_i\cdot \vec{n}_j)^2 \, dA.
\end{equation}\label{eq:nematic}

Here, $w$ represents the strength of the direct interaction constant, and the summation runs over all protein-protein pairs. The sum of $E_1$ and $E_2$ is minimized numerically, while $E_2$ is given by either isotropic (Eq. \ref{eq:eq3}) or nematic binding  (Eq. \ref{eq:eq6}). Isotropic interaction reflects the tendency of CMCs to self-aggregate, while nematic interaction also describes the tendency to align CMCs' principal curvatures. All CMCs in this work are either arc or saddle-shaped (see Figure 1.1(a))

The details of the Monte-Carlo procedure are given in Appendix \ref{app:procedure} and the details of the mesh generation and finding the principal curvatures are given in in Appendix \ref{app:code details}. 

\subsection{Reduced volume}

A volume constraint on the vesicle shapes can be implemented by adding an energy term: 
\begin{equation}\label{eq:eq7}
    E_3 = \frac{k_v}{2}\left(\bar{v} - \frac{6\sqrt{\pi} V}{A^{1.5}}\right)^2
\end{equation}
Here, the $\bar{v}$ is the target reduced volume that can vary from 0 to 1. A sphere has a reduced volume of 1. $A$ and $V$ are the area and the volume of the vesicle, respectively. The volume modulus $k_v$ was determined to be of the same order of magnitude as the bending and bonding terms.  The sum of $E_1$, $E_2$ and $E_3$ is minimized numerically. 

\subsection{Order parameter}

We want to quantify the degree of order between neighboring anisotropic CMCs. For this reason we can define the nematic order of the inclusion at vertex $i$ by
\begin{equation}\label{eq:eq8}
    S_i = \frac{1}{2}(3\cos^2{\theta}-1),
\end{equation}
where $\theta$ measures the angle between directors (or principal curvature $C_{1m}$) of neighboring CMCs on the membrane. The more aligned the neighboring CMCs are, the more this value approaches unity. In our numerical approach it is calculated by a Python script: for every CMC occupied vertex, it identifies neighboring vertices and uses \ref{eq:eq8} through a dot product for the ones which are also occupied. An arithmetic average of these is then calculated and recorded in a value $S_i$. The average nematic order across the entire membrane is then
\begin{equation}\label{eq:eq9}
    \langle S \rangle = \frac{1}{N}\sum_{i=1}^N S_i.
\end{equation}
\subsection{Mean cluster size and the gyration tensor}

We can define a mean cluster size $\langle N \rangle$ for simulations where the vesicles are not fully covered by CMCs. If we index all clusters so that $i$ has $N_i$ vertices, $\langle N \rangle$ is calculated as: 
\begin{equation}\label{eq:eq10}
    \langle N \rangle = \frac{\sum_{i}N_i}{\sum_{i} 1} = \frac{N_{\rm{vertex}}}{N_{\rm{clusters}}}. 
\end{equation}
The mean cluster size is not a reliable way to differentiate between phases. While all phases contain large clusters, their structure and organization vary significantly. This measure is heavily influenced by the total number of clusters, giving too much weight to small, single-vertex clusters. As a result, it introduces too much noise to effectively separate most phases.

To clearly distinguish phases in which CMCs form large condensed clusters, we instead rely on morphological measures. The shape of the vesicle is captured through the eigenvalues of the gyration tensor, $\lambda^2_i$. The gyration tensor \cite{theodorou1985shape} is defined as the average over all the vertices, with respect to the center of mass, similar to the moment of inertia tensor for equal-mass vertices: 
\begin{equation}\label{eq:eq11}
    R_{G_{ij}} = \langle r_i r_j \rangle = \dfrac{1}{N} \sum_{\rm{vertices}} \begin{pmatrix}
x^2 & xy & xz\\
xy & y^2 & yz\\
xz & yz & z^2
\end{pmatrix}.
\end{equation}
This can be visualized by a unique ellipsoid which has the same gyration tensor
\begin{equation}\label{eq:eq12}
\textit{\textbf{x}}^{T}\textit{\textbf{R}}^{-1}_{\textit{\textbf{G}}}\textit{\textbf{x}}=\dfrac{(\textit{\textbf{x}} \cdot \textit{\textbf{e}}_1)^2}{\lambda^2_1}+\dfrac{(\textit{\textbf{x}} \cdot \textit{\textbf{e}}_2)^2}{\lambda^2_2}+\dfrac{(\textit{\textbf{x}} \cdot \textit{\textbf{e}}_3)^2}{\lambda^2_3}=3.
\end{equation}
The eigenvectors $(\textit{\textbf{e}}_i)$ of the gyration tensor are the directions of the semi-axes of the equivalent ellipsoid and the
eigenvalues are their length squared divided by 3, ordered by their size: $\lambda^2_1 \leq \lambda^2_2 \leq \lambda^2_3$. 

\section{Materials and methods}

The simulations were run using trisurf-ng \cite{fosnaric2019theoretical} with a tape file modified from the available default with the different physical parameters and additional simulation parameters of nshell=20 ($N=2002$), mcsweeps=100,000, iterations=500 (MC steps). The fraction of CMCs on the membrane is set at the beginning of the simulation and is kept constant $\rho=N_{\rm{CMC}}/N$. Simulations were run in parallel, each on a separate core of a 20-core PC, with one needing usually 16 hours to finish. The resulting VTU files were viewed and prepared for figures publication in ParaView, but further analysis and graph generation were done by separate python scripts. The bending rigidity of the simulated membranes is 20 $k_B T$ throughout. The curvature of the CMCs throughout the paper is given in units of  $1/l_{min}$, which means that spontaneous curvature $C_{1m}=0.25$ corresponds to the curvature of the sphere with radius $4  l_{min}$. Similarly, all energy values throughout the paper are given in units of $k_B T$, where $k_B$ is the Boltzmann constant and $T$ absolute temperature.

\section{Results}\label{sec:result}
 
\subsection{Fully occupied vesicles with no volume constraints}

We first investigate the steady-state shapes of vesicles completely covered with CMCs ($\rho=1$) without imposing volume constraints. 

\subsubsection{Arc-shaped CMCs promote the formation of elongated cylindrical vesicles ($H_m=D_m>0$)}

\begin{figure}[t]
    \centering
    \includegraphics[width=\textwidth]{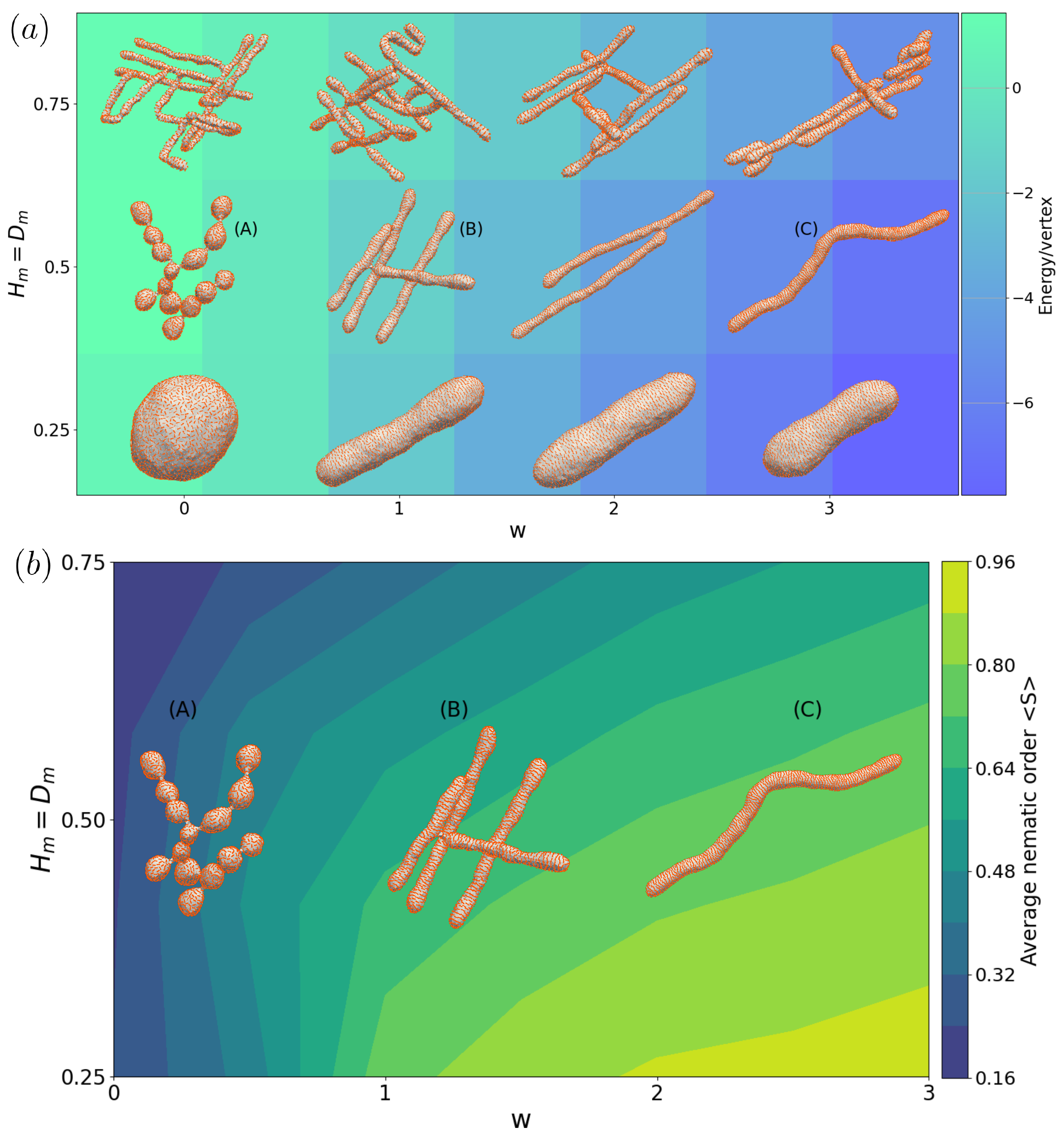}
    \caption{(a) Vesicles fully covered ($\rho=1$) with arc-shaped CMCs as function of the nematic interaction energy $w$ and intrinsic mean curvature $H_m$. Steady-state shapes are generally all cylinders (shapes (B), (C)). The pearling steady-state shapes in (A) arise as a consequence of neighboring CMCs assuming random orientations. Even in the absence of nematic interactions between neighboring CMCs, the membrane conforms to their spontaneous curvature. The heat map gives the total energy of each shape per vertex ($E_1 + E_2$ from Eqs. (\ref{eq:eq2}) and (\ref{eq:eq6})), which decreases with $w$. (b) A heat map of the average nematic order $\langle S \rangle$ reveals that it increases with $w$, but is most pronounced when CMCs are less curved. The shapes characterized by junctions or necks have a low degree of nematic order (the pearling phase), as seen in the close up of Figure \ref{fig:fig5}(A) . Orange lines show the direction of the principal CMC curvature $C_{1m}$.}
    \label{fig:fig2}
\end{figure}

The binding energy-spontaneous curvature phase diagram for vesicles fully covered with arc-shaped CMCs, without volume constraints, is shown in Figure \ref{fig:fig2}. Direct nematic interaction energy $w$ and intrinsic curvature of CMCs increase along the $x$ and $y$  axes, respectively.  Generally, equilibrium shapes are cylindrical with the radius of the cylinders determined by the curvature of the CMCs. In the case of no interaction $w=0$ and low spontaneous curvature, (bottom left corner of Figure \ref{fig:fig2}), all CMCs have random orientations, resulting in a sphere. The heat map of Figure \ref{fig:fig2}  shows the energy density of each shape. When the neighbor interaction is negligible ($w=0$), the energy is predictably positive due to  membrane bending (the leftmost column of  Figure \ref{fig:fig2}), while an increase in $w$ increases the nematic order and lowers the overall energy.   

\begin{figure}[t]
    \centering
    \includegraphics[width=1.1\textwidth]{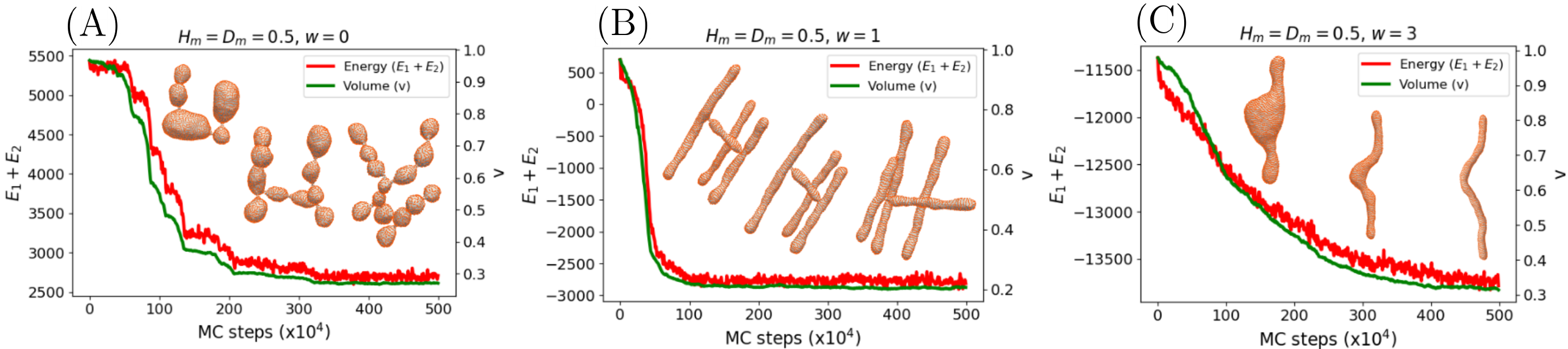}
    \caption{Convergence of bending and interaction energy (Eqs. \ref{eq:eq2} and \ref{eq:eq6}) for the evolution of shapes shown in Figure \ref{fig:fig2}. Convergence is achieved within 500 MC steps, which was taken as a default number of MC steps for each simulation.}
    \label{fig:fig3}
\end{figure}

\begin{figure}[t]
    \centering
    \includegraphics[width=1.1\textwidth]{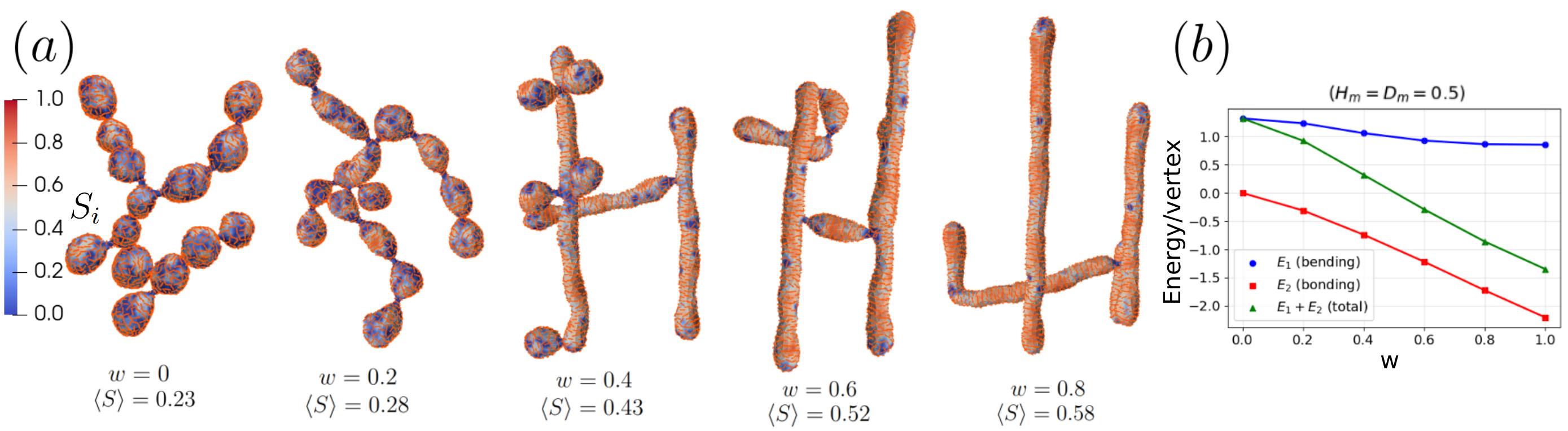}
    \caption{As nematic strength $w$ increases from zero, the steady-state shapes go from pearls to cylinders (a). This transition is marked by a steady decrease of not only bending, but also total energy (b). The parameters are the same as in panel (A) of Figure \ref{fig:fig2}(a).}
    \label{fig:fig4}
\end{figure}

\begin{figure}[t]
    \centering
    \includegraphics[width=1.1\textwidth]{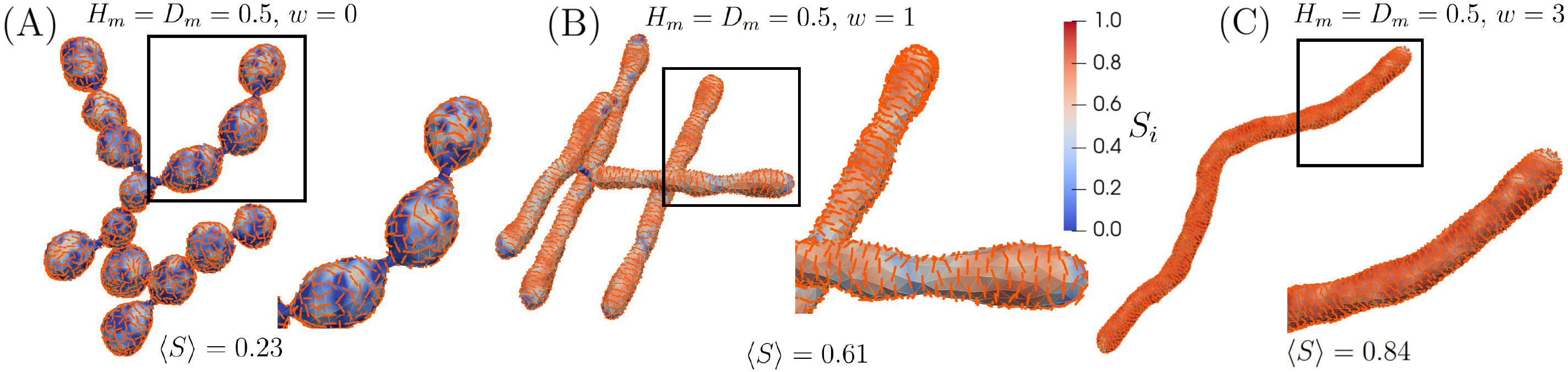}
    \caption{Local nematic order (Eq. \ref{eq:eq8}) and close up of the shapes shown in Figure \ref{fig:fig2}. Also shown is the average nematic order $\langle S \rangle$ (Eq. \ref{eq:eq8}).}
    \label{fig:fig5}
\end{figure}

Figure \ref{fig:fig3} shows the convergence of the bending energy and volume for three shapes of Figure \ref{fig:fig2}, as function of MC steps. We see that convergence is achieved in all cases, but at different rates. At $H_m=0.5$ and $w=0$, the vesicle forms a shape composed of smaller spheres joined together to form a series of pearls connected by thin necks, the pearling phase (Figure \ref{fig:fig2}(A)). In the neck regions, $S_i$ is low. An gradual increase in $w$ from zero leads to fewer necks as the neighboring CMCs align to form adjoined smaller cylinders with large nematic order $\langle S\rangle$  (Figure \ref{fig:fig4}). Figure \ref{fig:fig5} shows close-ups of the steady-state shapes in Figure \ref{fig:fig2} with the color map indicating the local nematic order parameter $S_i$ (Eq. \ref{eq:eq8}) and the average nematic order of the shapes $\langle S \rangle$ (Eq. \ref{eq:eq9}). 
 
\begin{figure}[ht]
    \centering
    \includegraphics[width=0.8\textwidth]{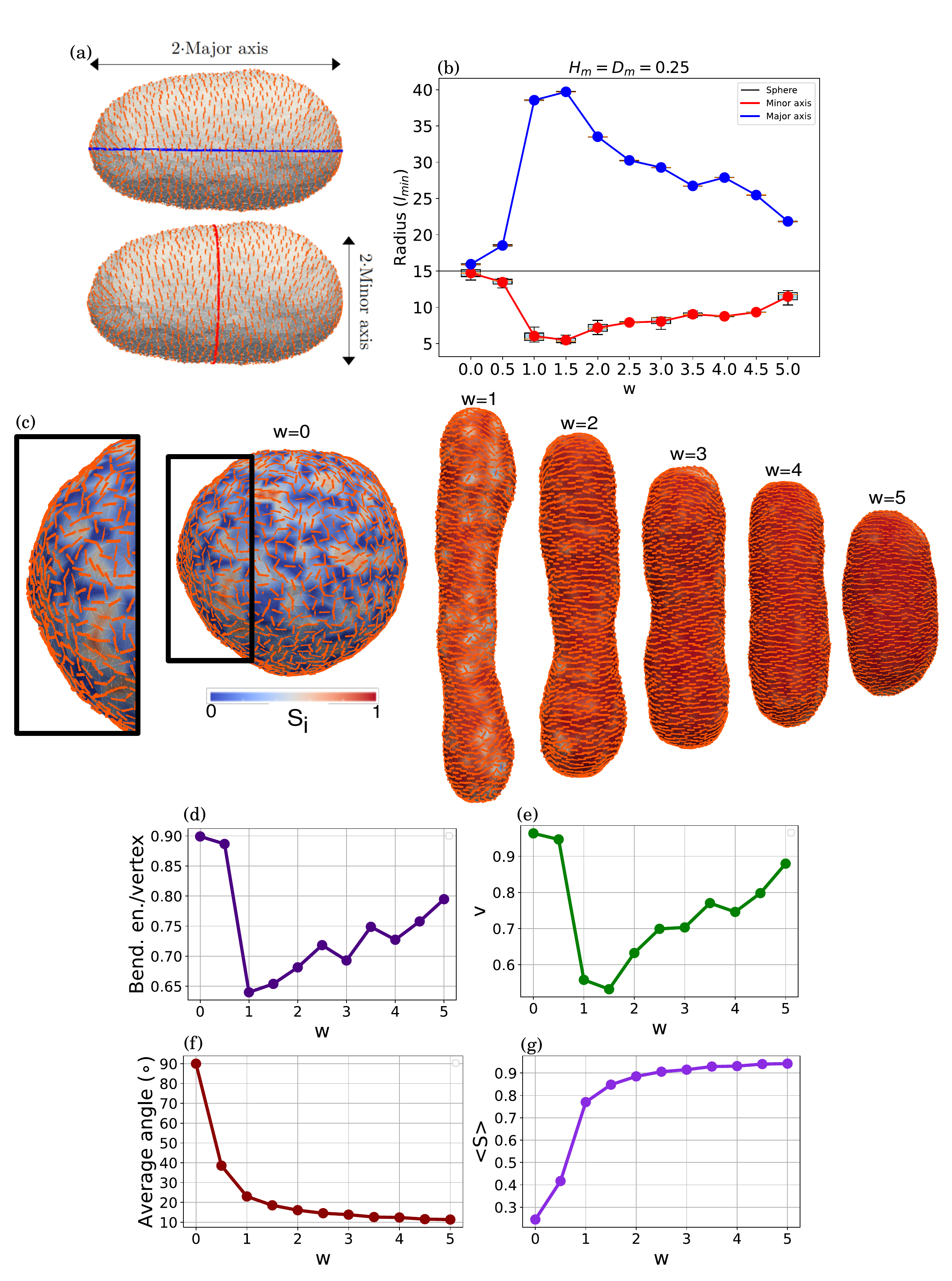}
    \caption{(a) Steady-state equilibrium shapes that are fully covered ($\rho=1$, $H_m=D_m=0.25$) by arc-shaped CMCs result in prolate phases for $w>0$. For a prolate, the major and minor axes can be determined. At $w=0$, the neighboring CMCs do not interact and have no effect on the orientations of their neighbors, resulting in a spherical shape with both major and minor axes having the same length (b, c). Increasing $w$ forces the neighboring CMCs to align with each other, which leads to elongation of the vesicles (b, c). The bending energy per vertex has a minimum at $w=1$, before starting to increase due to stronger bonding (d). The elongation of the vesicles is most pronounced when $w=1.5$ and $v=0.56$ (e). The average angle between neighboring vertices decreases as function of $w$ (f), while the inclusion average order increases with $w$ (g). When $w$ is larger than $1.5$, the vesicles become less elongated and tend towards the spherical shape for large values of $w$ (c).}
    \label{fig:fig6}
\end{figure}

The bottom row of Figure \ref{fig:fig2}(a) reveals that for spontaneous curvature $H_m=D_m=0.25$ and increasing $w$, the steady-state shapes change from a sphere to a prolate state of different radii, in a nonmonotonous manner. This process is analyzed in greater detail in Figure \ref{fig:fig6}. Each prolate shape is approximately an ellipsoid that is characterized by two axes along its symmetrical planes. As the area is kept constant, an increase in $w$ from zero results in an elongation of the axis in one direction and a narrowing in another (Figure \ref{fig:fig6}(a, b, c)). As $w$ increases from zero, the CMCs begin to align with each other, transforming the initial spherical shape into an elongated prolate phase (Figure \ref{fig:fig6}(c)), resulting in a decrease of bending energy and reduced volume (Figure \ref{fig:fig6}(d, e)). The in-plane rotation accompanying this change is reflected in the decrease in the average angle between neighboring CMCs (Figure \ref{fig:fig6}(f)). $\langle S \rangle$ monotonically increases with increasing $w$ (Figure \ref{fig:fig6}(g)). Since the CMCs do not perfectly align perpendicular to the axis of the tube, the tube has a radius that is smaller than the spontaneous radius of curvature of the CMC. As $w$ increases, the CMCs align more perfectly with respect to each other, and more perpendicular to the tube axis, which increases in radius until it equals the radius of the CMC for large $w$.

\subsubsection{Saddle-like CMCs promote global saddle shapes ($H_m=0$, $D_m>0$)}

\begin{figure}[ht]
    \centering
    \includegraphics[width=\textwidth]{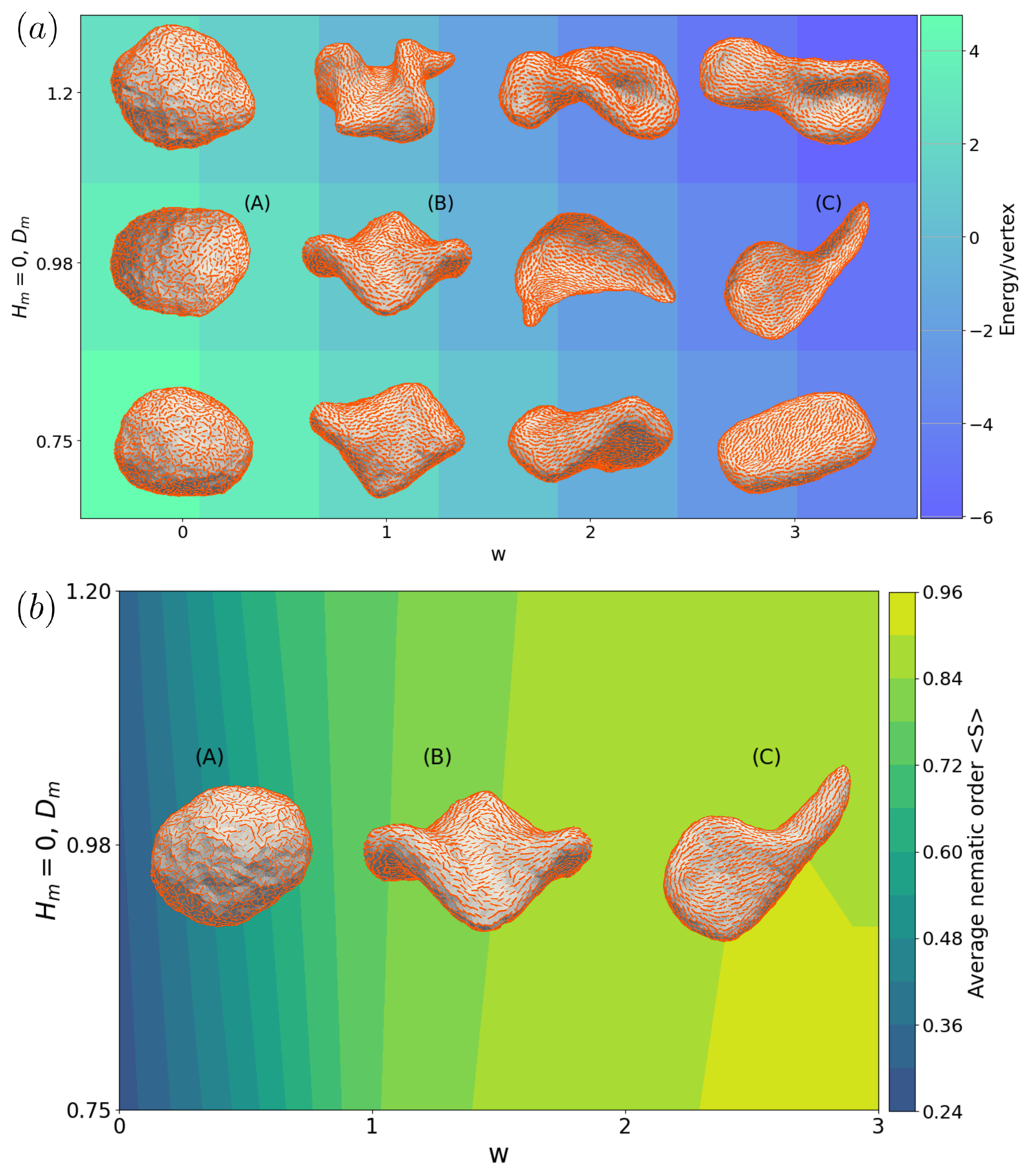}
    \caption{
(a) The curvature-binding strength phase diagram for steady-state shapes with saddle CMCs ($\rho=1$). The curvature of the hyperbolic paraboloid is given on the $y$-axis by $D_m$ ($H_m=0$) and $w$ on the $x$-axis. When saddle CMCs are more flat ($D_m=0.75$), the vesicles mirror this by exhibiting regions of flat membrane (top row), while more curved saddle CMCs ($D_m=1.2$) result in global shapes that mimic the inclusion's curvatures (bottom row). The energy per vertex is shown by the corresponding heatmap. (b) Average nematic order heatmap for the shapes in (a).}
    \label{fig:fig7}
\end{figure}

\begin{figure}[ht]
    \centering
    \includegraphics[width=\textwidth]{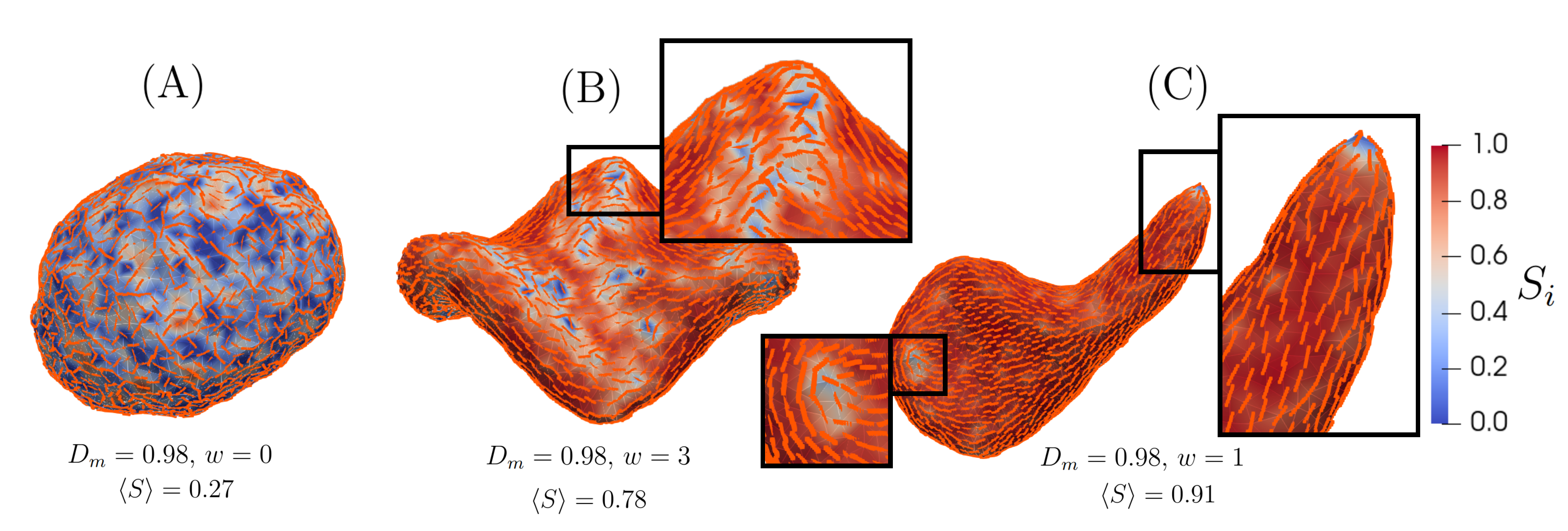}
    \caption{Nematic order for the steady-state shapes of vesicles covered by saddle-shaped CMC, shown in Figure \ref{fig:fig7}(b). The nematic defects characterized by minimal $S_i$ are shown in blue. When there is no interaction $w=0$, the steady-state is a sphere (a) and a bowtie-like morphology for $w=1$ (b). The inset shows the nematic defect field, which is not fully developed. (c) The nematic defect is most notable when $w=3$, at the tip of the elongated membrane. Its topological charge is +1, while there are two additional charges +1/2 on either side of the wider part (inset). }
    \label{fig:fig8}
\end{figure}

Saddle-like CMCs are inherently frustrated on the convex surface of the vesicle. For these CMCs, $H_m=0$ and $D_m>0$. The shape of the saddle of the CMCs is determined by the curvature of the hyperbolic paraboloid which is in turn defined by the deviatoric component $D_m$; a flat saddle corresponds to $D_m=0$, while curved saddles have non-zero values $D_m$. Independently varying $D_m$ and $w$ leads to the phase diagram shown in  Figure \ref{fig:fig7}. For non-interacting CMCs, $w=0$, vesicles remain spherical, as in the case of arc-shaped CMCs,  due to their random orientations. By contrast to arc-shaped CMCs, there is no spontaneous budding  even as the spontaneous curvature increases. As $w$ increases, the vesicles deform from the spherical shape. In the case of $D_m=0.75$, the vesicle develops globally flattened regions of the membrane with increasing $w$ (see bottom row of Figure \ref{fig:fig7}). In the case of most curved saddles ($D_m=1.2$), the vesicle features a bow-tie phase. (see top row of Figure \ref{fig:fig7}).  The heat map shows that the energy density decreases with the increase of $w$, consistent with observations for arc-shaped CMCs (Figure \ref{fig:fig2}). Figure \ref{fig:fig8} shows the local nematic order for the steady-state shapes in Figure \ref{fig:fig7}(b), with saddle-like CMCs and an increasing interaction $w$. Larger $\langle S \rangle$ leads to more localized nematic defects, shown in blue.

\subsection{Partly covered vesicles with volume constraints}

We next investigate the steady-state shapes of vesicles that are only partially covered with CMCs, while also imposing a fixed volume constraint.

\subsubsection{Phase diagram for vesicles half covered with arc-shaped CMCs  ($H_m=D_m=0.25$)}

\begin{figure}[ht]
    \centering
    \includegraphics[width=\textwidth]{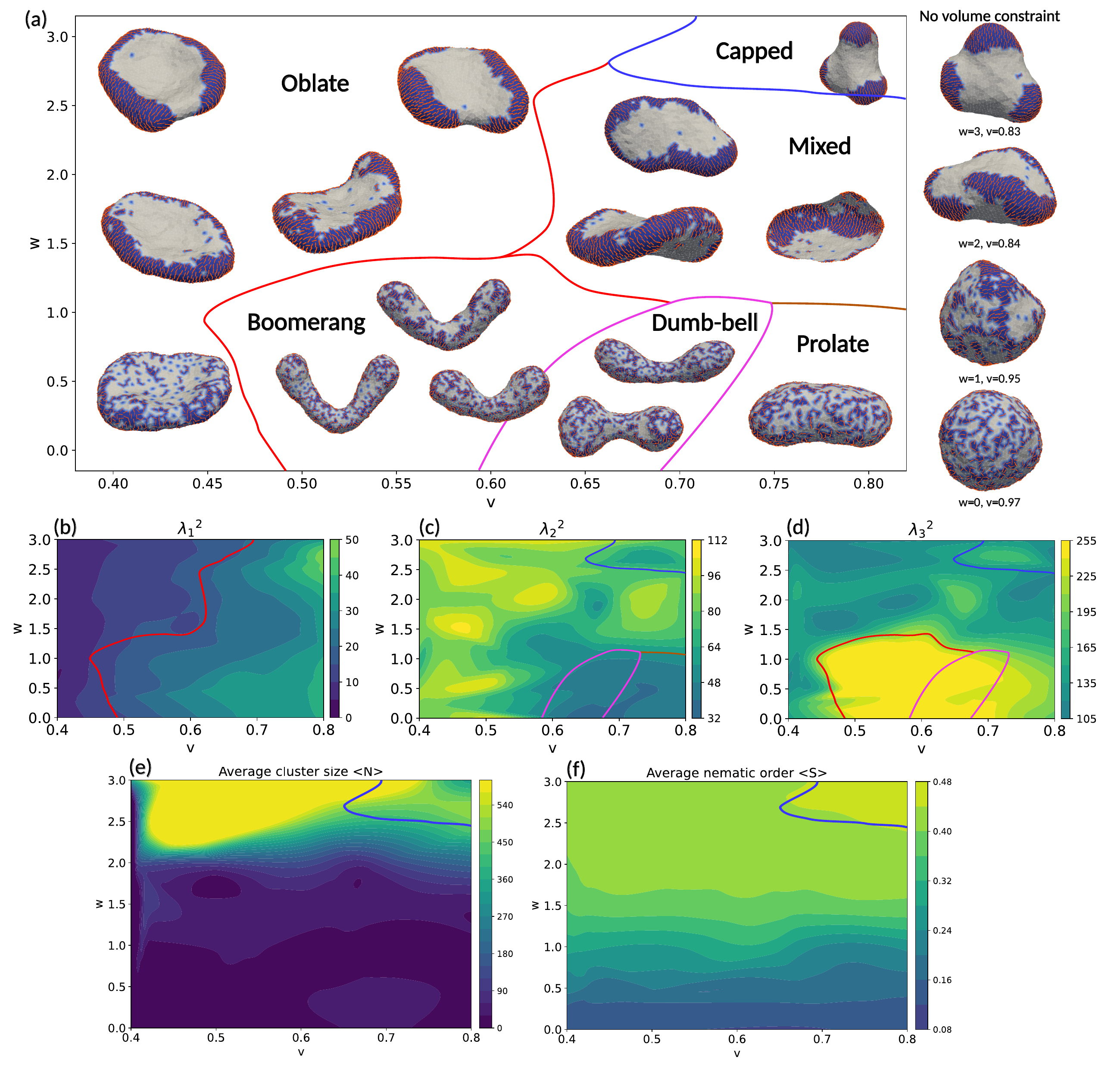}
    \caption{Reduced volume-binding strength plane for arc-shaped CMCs. (a) Phase diagram as function of $v$ and $w$ for $\rho=0.5$ CMC concentration for $H_m = D_m = 0.25$. The different phases are indicated by their names and a typical snapshot of the equilibrium shape is shown. The transition lines between the phases were drawn according to the measures shown in the bottom panels. The oblate phase is separated by the small eigenvalue $\lambda^2_1$, as explained in the main text (b). The prolate phase is roughly indicated by a small intermediate eigenvalue $\lambda^2_2$ (c). The prolate phase is indicated with largest eigenvalue $\lambda^2_3$, but the boomerang and dumb-bell phases' transition is difficult to determine from $\lambda^2_i$ (d). The average cluster size heatmap shows an approximate separation between the oblate and the capped phases. When $v$ is increased above 0.7, the CMC patch on the rim of the oblate phase segregates into distinct CMC patches. This also results in the capped phase having the largest average nematic order $\langle S \rangle$ (f). The column right of panel (a) shows that a decrease in $v$ is slight with no volume constraints and increasing $w$, resulting in a transition that goes straight from spherical to capped phases.}
    \label{fig:fig9}
\end{figure}

First we explore the phase diagram of steady-state equilibrium shapes for $\rho=0.5, H_m=D_m=0.25$ in the $v-w$ plane (Figure \ref{fig:fig9}). We identify several phases: oblate, prolate, and capped, with the boomerang and dumb-bell phases as subsets of the prolate phase, and the mixed phase as a subset of the oblate phase (see Figure \ref{fig:fig9}(a)). The phases can be roughly discerned by the eigenvalues $\lambda^2_i$ of the gyration tensor (Eq. \ref{eq:eq11}), shown in Figure \ref{fig:fig9}(b-d).  The first eigenvalue $\lambda^2_1$ measures the thinness and is low for oblate shapes. The second eigenvalue $\lambda^2_2$ is large for the oblate shapes (and roughly equal to the largest eigenvalue $\lambda^2_2 \approx \lambda^2_{3}$), but is minimized for elongated prolate shapes (and roughly equal to the largest eigenvalue $\lambda^2_2 \approx \lambda^2_{1}$). The third eigenvalue $\lambda^2_{3}$ is largest for shapes that are elongated along one principal axis and characterizes the prolate phase. Since the separation between boomerang, dumb-bell and prolate phase is more qualitative than quantitative, as they are all marked by an elongated axis, the heatmap for $\lambda^2_{3}$ shows highest values for all three.  Figure \ref{fig:fig9}(e) shows the average cluster size $\langle N \rangle$ (Eq. \ref{eq:eq10}).  The oblate phase is characterized by CMC clustering on the rim to form discs (as in the case of isotropic inclusions \cite{ravid2023theoretical}). An increase in $v$ results in the rim of CMCs not closing up entirely, while still keeping their approximate oblate shape. Prolate shapes are limited to values of $w \leq 1$ and contain two sub-phases: the boomerang and the dumb-bell. The oblate phase transitions into the mixed phase approximately above $v=0.65$, and the prolate below $w=1.5$. The mixed phase is characterized by two effects: stronger CMC clustering due to high $w$ and reduced flatness due to large $v$. The average nematic order $\langle S \rangle$ (Eq. \ref{eq:eq9})  is shown in Figure \ref{fig:fig9}(f). Predictably, $\langle S \rangle$ increases with increasing CMC alignment and $w$. The capped phase is located at high $v$ and $w$. Here, the clusters form separate patches, sometimes with a pronounced "cap" at the tips of shape protrusions. The patch separation results in a lower $\langle N \rangle$, but better nematic order of CMCs with higher $\langle S \rangle$, both discernible on the heatmaps in Figure \ref{fig:fig9}(e-f). Below $w=1$ the average cluster size is $\leq 100$ (Figure \ref{fig:fig9}e) and the CMC distribution is dominated by mixing entropy and smaller $\langle N \rangle$.  A comparison of shapes with the same interaction $w$ but no volume constraint is shown on the right of Figure \ref{fig:fig9}. At $w=0$ and 1, the shape is roughly spherical, then changes to the mixed and capped phases at $w=2$ and $w=3$, respectively.  

\subsubsection{Arc-shaped CMCs ($H_m=D_m=0.25$) accelerate the oblate-prolate shape transition in comparison to bare-membrane vesicles}

\begin{figure}[t]
    \centering
    \includegraphics[width=\textwidth]{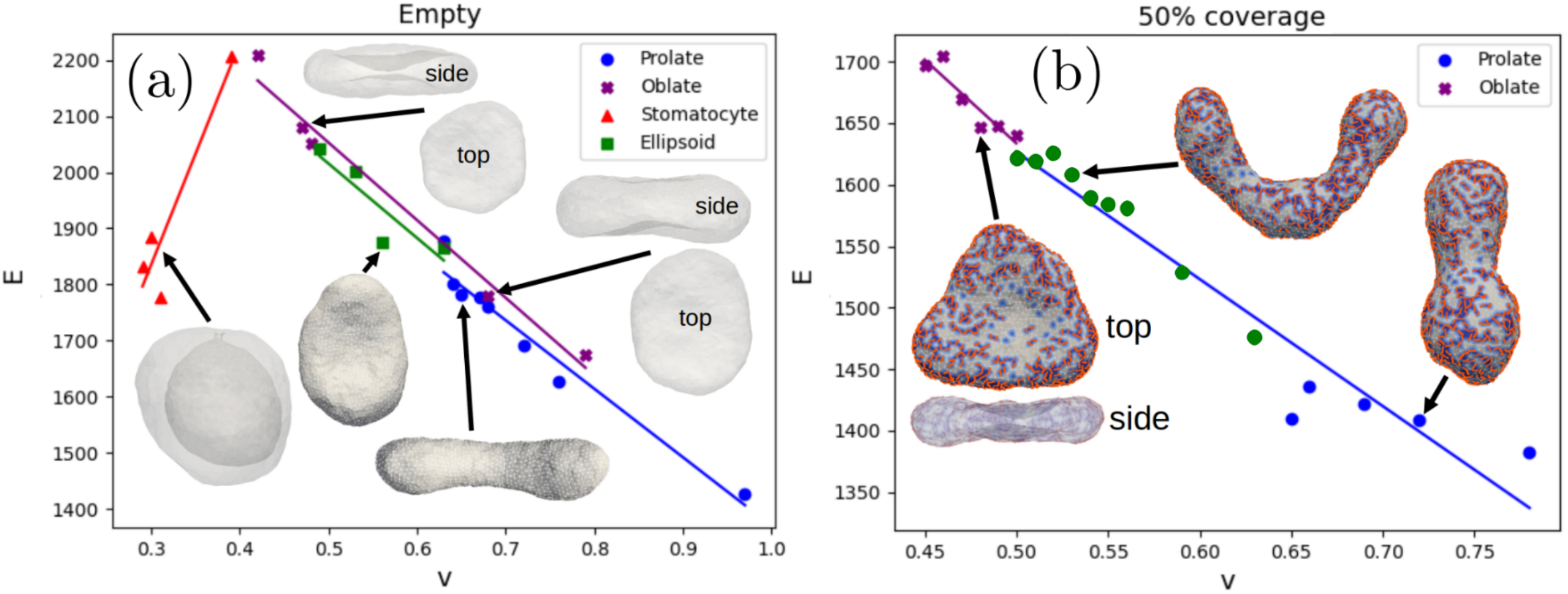}
    \caption{(a) The phase diagram in the space $v$-$E$ for vesicles with no CMCs shows a familiar sequence of morphologies with increasing $v$; stomatocytes, oblates and prolates. An interesting intermediate shape transition (ellipsoid) is observed between oblate and prolate shapes, marked in green, which was reported in literature  \cite{kralj1993existence} . (b) The phase diagram in the $v$-$E$ space for arc-shaped CMCs ($\rho=0.5, H_m=D_m=0.25$, $w=0$). In comparison to empty vesicles, there is no stomatocyte phase, but an intermediate boomerang phase (shown in green) which is a subset of a prolate phase. The transition to prolates is accelerated and happens at around $v=0.5$, possibly due to entropic mixing of the CMCs, which leads to their migration away from the rim of the oblate vesicles.}
    \label{fig:fig_10}
\end{figure}

We plot the $v-E$ diagram of vesicles at steady-state (Figure \ref{fig:fig_10}(a)), that contain no CMCs, in agreement with existing literature \cite{seifert1991shape}. Starting from $v=0.3$ and increasing $v$ the steady-state shapes follow a familiar pattern of the stomatocyte, oblate and prolate phases (Figure \ref{fig:fig_10}(a)). The phases are discerned from the eigenvalues of the gyration tensor $\lambda^2_i$. There is overlap where the oblate and prolate phases coexist, roughly in the range $v=0.6$-$0.8$. However, in this region, $E$ is lower for prolates, which extend to $v=1$.  We next examine how the steady-state vesicle shapes change when they are half covered with arc-shaped CMCs ($\rho=0.5$), and how the $v-E$ phase diagram compares to the bare membrane vesicles. The results are shown in Figure \ref{fig:fig_10}(b). We now find that the oblate phase has been pushed to exist only below $v \approx 0.5$, while the prolate phase is now extending over a larger range of reduced volume (compared to Figure \ref{fig:fig_10}(a)).

\begin{figure}[t]
    \centering
    \includegraphics[width=0.5\textwidth]{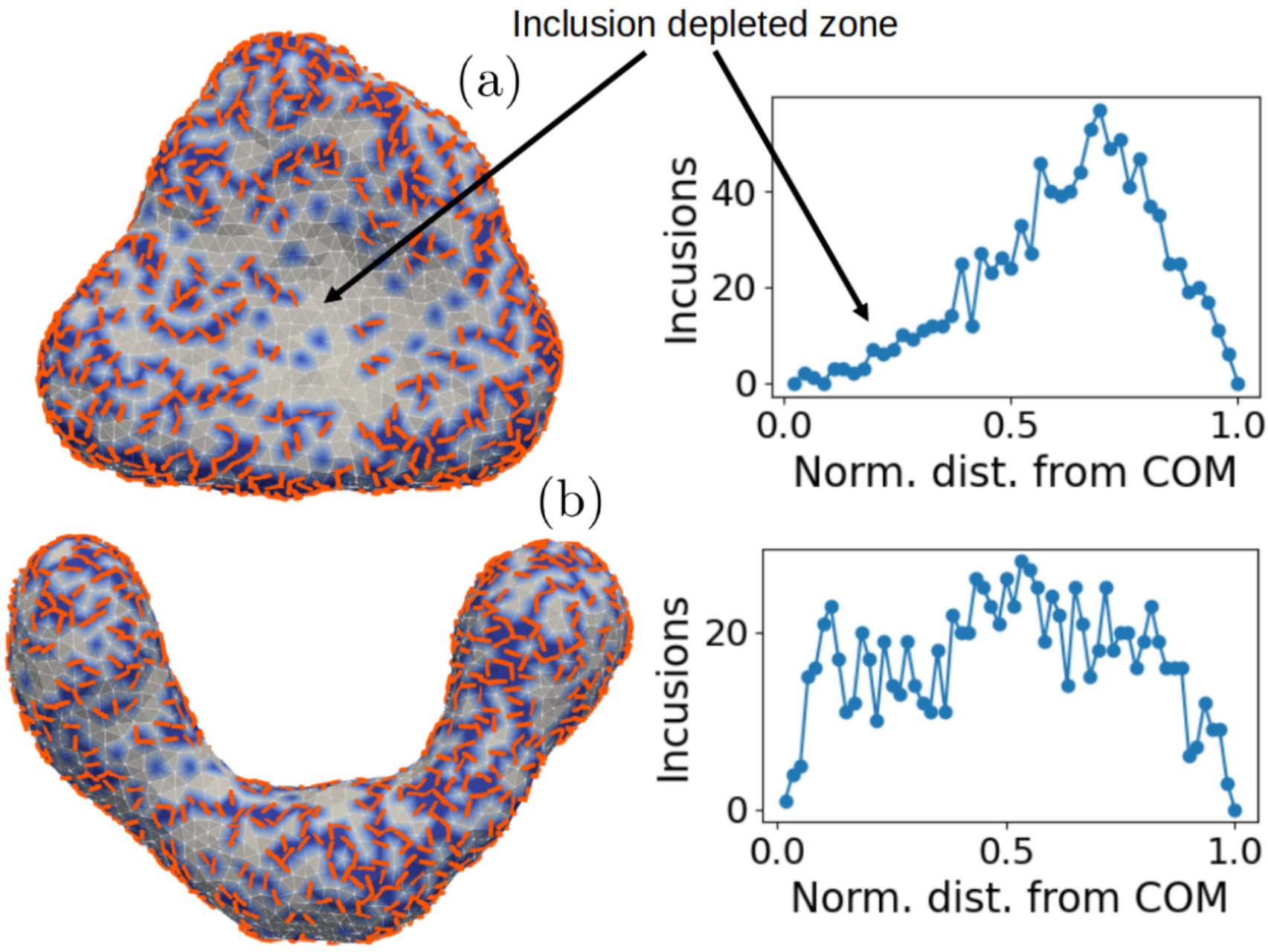}
    \caption{The oblate-prolate phase transition includes a shift of the inclusion distribution measured from the center of mass (COM). Oblate vesicles have most of their CMCs on the rim (a) while in prolate vesicles, these are more evenly distributed (b).}
    \label{fig:depletion}
\end{figure}

The smaller regime of oblate stability is driven by the lower mixing entropy of the CMCs on the oblate compared to the prolate shape. The oblate phase has two flat sides where the arc-shaped CMCs face a bending energy penalty, which is reduced when the CMCs accumulate on the curved rim. This is not the case for prolates, which have more uniform curvature. We can observe the effect of mixing entropy by plotting the distribution of CMCs as function of distance from the center of mass (COM) of the vesicle (Figure \ref{fig:depletion}). When $w$ is increased, the CMC distribution is more pronounced on the rim of the oblate shapes, lowering their entropy (see Supplementary Information \ref{app:supp2}).


\subsubsection{Phase diagram for vesicles half covered with saddle-like CMCs  ($H_m=0$, $D_m=0.98$)}

\begin{figure}[ht]
    \centering
    \includegraphics[width=\textwidth]{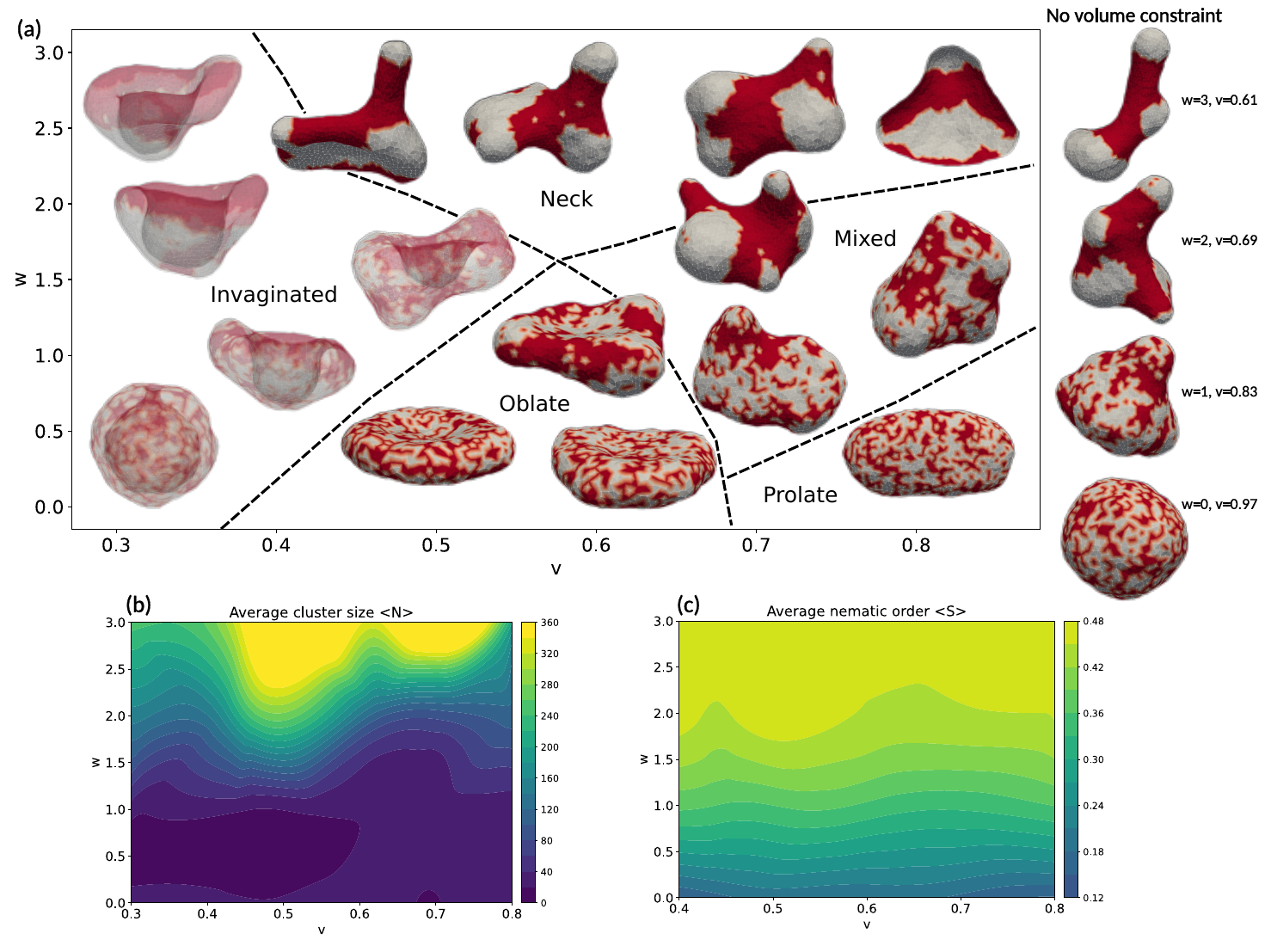}
    \caption{Reduced volume-binding strength plane for saddle-like CMCs. (a) Phase diagram as function of $v$ and $w$ with $\rho=0.5$ concentration for $H_m = 0$ and $D_m = 0.98$. Invaginated stady-state shapes are common under $v=0.5$ with the majority of CMCs found on the invaginated rim. The oblate shapes are limited to low $w$ and regions approximately between $v=0.5-0.6$. An elongated neck-like formation is typical at $w=3$, but the elongation becomes less pronounced at higher $v$. In comparison, steady-state shapes with no volume constraints are shown right of panel (a). The average cluster size heatmap shows largest clusters at $w=3$, approximately where neck formation takes place (b). Average nematic order gradually increases with increasing $w$ (c). The transitions between the phases are less defined than in the case of arc-shaped CMCs, with oblate and invaginated phases often having very similar equilibrium energies.}
    \label{fig:fig12}
\end{figure}

We plot the $v-w$ phase space for the steady-state shapes of vesicles that are half-covered by saddle-like CMCs with  $H_m=0$ and $D_m=0.98$ (Figure \ref{fig:fig12}). Saddle CMCs with negative Gaussian curvature are inherently frustrated on convex membranes. The phase diagram features a high degree of metastability, resulting in approximate phases that often coexist for the same $v$ and $w$. This is also evident in the heatmaps of $\lambda^2_i$, which cannot reliably distinguish between phases (not shown). The poor reproducibility of these transitions warrants further investigation but lies beyond the scope of this paper.  The simulations reveal four phases with ambiguous transition boundaries: the invaginated, oblate, neck, mixed, and prolate phases. 

The invaginated phase is found at $v<0.4$, with most CMCs localized on the saddle-curved rim of the invagination. The oblate phase appears for $w<1$ and $v=0.5$-$0.6$. The prolate phase is restricted to $v=0.8$ and $w<0.5$. The neck phase exhibits slender cylindrical protrusions with nematically ordered CMCs, flanked by convex, CMC-free regions. Notably, this phase persists even without volume constraints at $w=2$-$3$ (see Figure \ref{fig:fig12}(a), right column). Heatmaps of the mean cluster size $\langle N \rangle$ and mean nematic order parameter $\langle S \rangle$ (Figure \ref{fig:fig12}(b-c)) show a monotonic increase with $w$, where the neck phase corresponds to elevated $\langle N \rangle$.  

\begin{figure}[h!]
    \centering
    \includegraphics[width=1\textwidth]{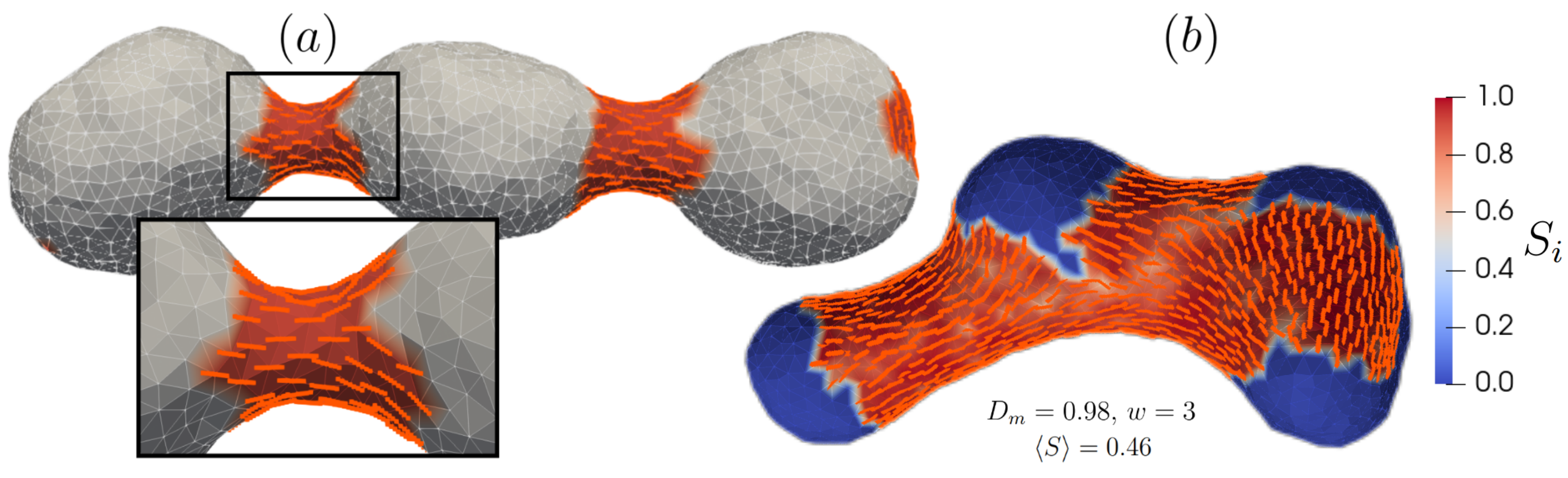}
    \caption{(a) Anisotropic saddle CMCs ($H_m=0$, $D_m=0.98$, $\rho=0.16$, $w=3$, $v=0.54$) form necks between empty convex membrane regions. Orange lines show the direction of the principal CMC curvature $C_{1m}$. (b) The nematic order heatmap and orientation of saddle-like CMCs for the case of $H_m=0$, $D_m=0.98$, $\rho=0.5$, $w=3$, $v=0.61$.}
    \label{fig:fig13}
\end{figure}

At a lower CMC density ($\rho=0.16$), steady-state shapes display CMC aggregation in neck regions between convex membrane segments, resembling a pearling phase  (Figure \ref{fig:fig13}(a)). For $\rho=0.5$ and $w=3$, saddle CMCs exhibit strong nematic alignment (Figure \ref{fig:fig13}(b)).

\subsubsection{Analysis of topological defects}

A topological defect (TD) on a surface arises when the order parameter $S_i$ of the inclusions cannot be smoothly defined everywhere, leading to singular points or lines where the order parameter abruptly changes. TDs are characterized by their discrete topological charge, an additive and conserved quantity \cite{mermin1979topological}. TDs with like charges repel each other, whereas those with opposite charges attract. Given these similarities to electrostatic interactions, some studies \cite{mesarec2016effective, bowick2004curvature, vitelli2004anomalous} suggest an electrostatic analogy to describe the interplay between Gaussian curvature and TD configurations. TDs with positive (negative) value of topological charge are referred to as defects (antidefects) \cite{mesarec2023coupling,mesarec2016effective}. On 2D surfaces, the topological charge corresponds to the winding number of the nematic director field \cite{mermin1979topological, lavrentovich1998topological}.  According to the Gauss-Bonnet and Poincaré-Hopf theorems \cite{poincare1885courbes, kamien2002geometry}, TDs must exist in all non-toroidal topologies. These theorems dictate that the sum of all topological charges (the total winding number) equals 2 for 2D surfaces of spherical topology. In nematic ordering, the topological charge can be a multiple of half an integer.

It is well-established that Gaussian curvature strongly influences the location and number of TDs \cite{mesarec2016effective, bowick2004curvature, vitelli2004anomalous}. Regions with positive (negative) Gaussian curvature attract TDs of positive (negative) topological charge.  While TDs are energetically costly—leading systems to avoid them, often through the annihilation of defect-antidefect pairs into locally defect-free structures \cite{mesarec2016effective}—their presence is frequently unavoidable due to the system’s topology. 

\begin{figure}[h!]
    \centering
    \includegraphics[width=1\textwidth]{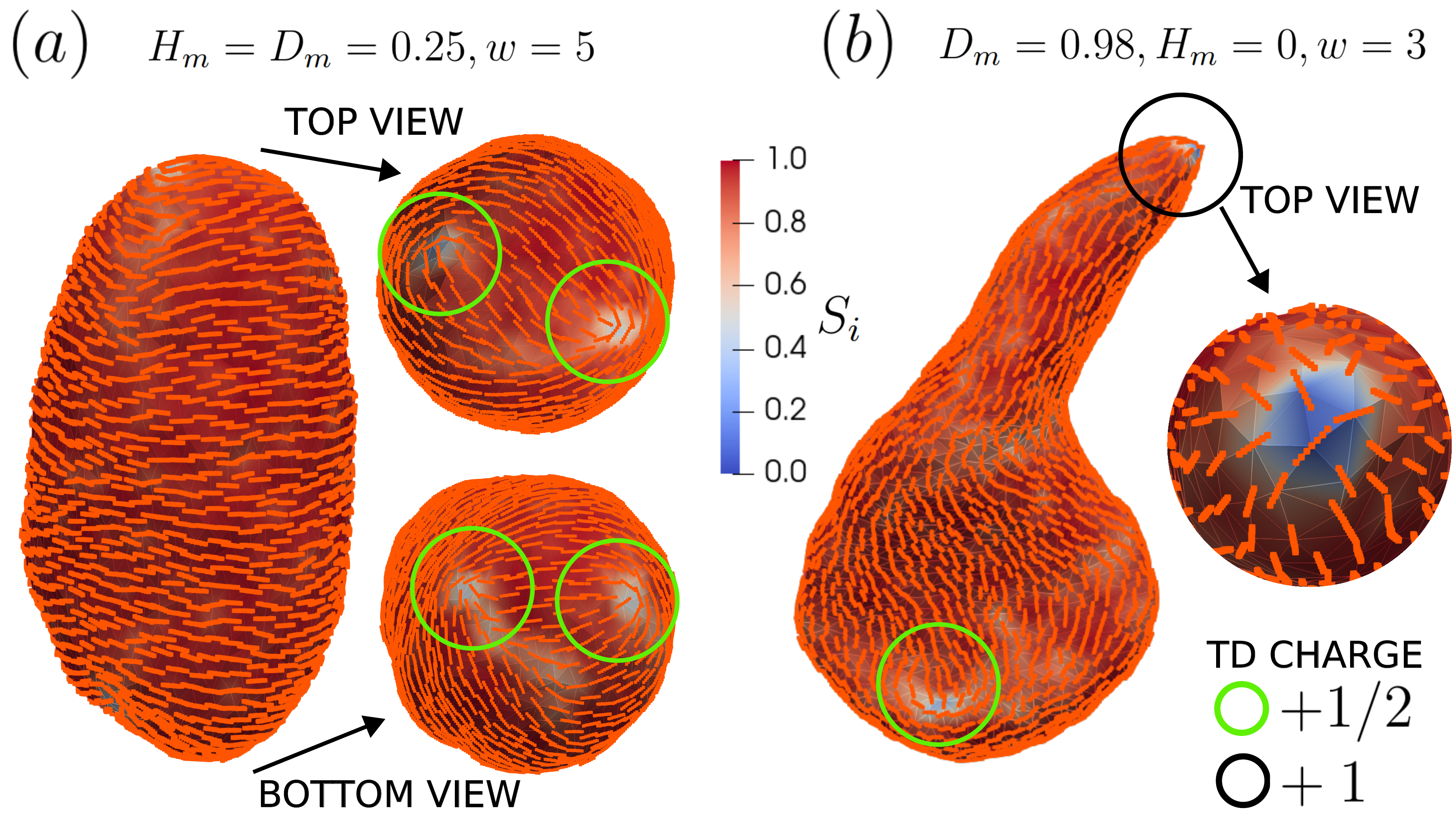}
    \caption{(a) For a fully covered equilibrium steady-state shapes with arc-shaped CMCs ($H_m=D_m=0.25, w=5$), the four defects of charge $+1/2$ are found on the cylindrical caps, where the Gaussian curvature is positive. (b) In the case of a membrane that is fully covered by saddle-like CMCs ($D_m=0.98$, $H_m=0$,  $w=3$), two $+1/2$ defects are found on either side of the bulbous base (the other defect is on the other side and not shown) and a $+1$ defect on the tip of the protrusion. The defects are characterized by a locally low nematic order $S_i$.}
    \label{fig:fig14}
\end{figure}

The occurrence of TDs is most apparent when CMCs cover the entire membrane surface and can be analyzed by studying $S_i$ and variations of $C_{1m}$ orientation (Figures \ref{fig:fig2}–\ref{fig:fig8}). In the case of vesicles fully covered with arc-shaped CMCs, the TDs of positive charge are normally found on the cap of the cylindrical shapes of the prolate phase where Gaussian curvature is positive.  An example of four defects with charge $+1/2$ can be seen in Figure \ref{fig:fig14}(a), which is a close-up of the TD found on the cap of the prolate phase in Figure \ref{fig:fig6}(c). Alternatively, when saddle-like CMCs fully cover the membrane, a single defect of charge $+1$ is found on the tip of the protrusion and two defects with charge $+1/2$ on either side of the base (Figure \ref{fig:fig14}(b)). They are positioned on opposite sides to maximize separation, consistent with the repulsion between like-charge defects. In both these cases, the topological charge sums up to $2$, in line with the Gauss-Bonnet and Poincaré-Hopf theorems.

The electrostatic analogy for TDs holds best when the intrinsic (direct interaction) term dominates \cite{bowick2004curvature, vitelli2004anomalous}. However, extrinsic \cite{napoli2012extrinsic, selinger2011monte} or deviatoric \cite{kralj2006quadrupolar, iglivc2005role} terms may also play a role, particularly on surface patches where principal curvatures differ. These terms can impose an external ordering field, favoring CMC orientations that best conform to the surface geometry. Consequently, intrinsic and deviatoric/extrinsic terms may introduce competing tendencies \cite{selinger2011monte}.

\begin{figure}[h!]
    \centering
    \includegraphics[width=1\textwidth]{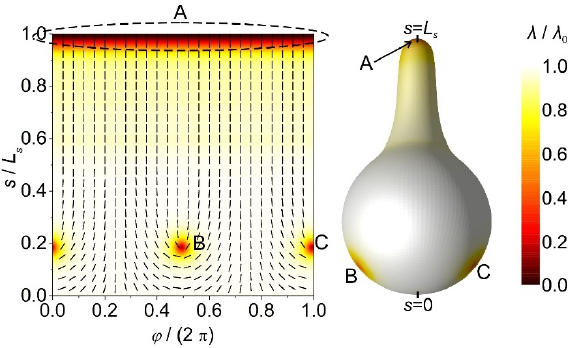}
    \caption{Equilibrium orientational ordering profile on a fixed axisymmetric 2D surface that is topologically equivalent to the shape presented in Figure \ref{fig:fig14}(b). The nematic ordering amplitude $\lambda$ is denoted by the color code, while the local orientation of molecules is presented by the lines in the $(\phi, s)$-plane, where $\phi$ is the azimuthal angle of the axisymmetric surface and $s$ the arc length of the profile curve characterizing the axisymmetric surface. The equilibrium nematic ordering amplitude is denoted by $\lambda_0$ and the total length of the profile curve by $L_s$. Topological defects are marked with capital letters. The topological charge of TD A, B and C is $+1$, $+1/2$ and $+1/2$, respectively. The surface is completely covered by saddle-like CMCs with $H_m=0$ and $D_m=0.98$. The details of the modeling used in this calculation are described in detail in \cite{mesarec2023coupling}. The only difference in the modeling is that we used the deviatoric term for 2D inclusions instead of the term for 1D inclusions, which was used in \cite{mesarec2023coupling}.}
    \label{fig:fig15}
\end{figure}

We compare the result shown in Figure \ref{fig:fig14}(b) to simulations of axisymmetric studies of closed membrane shapes with saddle-like CMCs and find good agreement with the equilibrium steady-state shape and with the position of defects (Figure \ref{fig:fig15}).

In our example (Figures \ref{fig:fig14}(b) and \ref{fig:fig15}), the deviatoric term (Eq. \ref{eq:eq2}) exerts its strongest influence along the tubular protrusion, where the difference in principal curvatures (i.e., the deviatoric curvature) is greatest. This term aligns the CMCs parallel to the protrusion, minimizing frustration. In contrast, the less deviatoric regions—particularly the nearly spherical base in Figure \ref{fig:fig15}—remain unaffected by this term, as the surface is isotropic (equal principal curvatures). The deviatoric term also has no effect at the highly curved, isotropic tip of the protrusion.

Since the TDs in Figures \ref{fig:fig14}(b) and \ref{fig:fig15} appear in regions of high Gaussian curvature, we conclude that intrinsic and deviatoric/extrinsic terms do not conflict in this case. The deviatoric term’s influence is evident along the tubular protrusion, where it enforces CMC alignment. In contrast, considering only the intrinsic term could yield configurations with CMCs tilted relative to the protrusion.

\section{Discussion}

Our results demonstrate that anisotropic curved membrane components (CMCs) significantly influence vesicle morphology through curvature sensing and nematic alignment. By focusing on the two simplest anisotropic CMCs — the arc- and saddle-shaped — we find a wider variety of steady-state phases compared to isotropic CMCs, where only convex budding phases were found across all CMC concentrations \cite{fovsnarivc2019,pandur2023surfactin}. By mapping the steady-state shapes of fully covered vesicles for both arc- and saddle-shaped CMCs, we found that in the case of arcs, there exists a competition between nematic ordering and curvature; when the former is absent, equilibrium steady-state shapes are a series of connected pearls (Figure \ref{fig:fig2}). Nematic binding tends to break up the pearling phase and results in smooth prolate phases, with the radius of the cylinder determined by the intrinsic curvature of the CMCs and entropy. With this, we have confirmed pearling using our non-axisymmetric numerical scheme and connected it to previous studies originating from analytic and theoretical considerations \cite{iglivc2005role,rueda2021gaussian}. Additionally, these findings align with experimental observations of BAR domain proteins (e.g., amphiphysin and IRSp53) stabilizing tubules or negatively curved membranes \cite{simunovic2015physics,johnson2024protein}. The pearling-to-cylinder transition at low nematic strength mirrors the budding and tubulation processes seen in cellular membranes, suggesting that weak interactions may suffice for initial curvature sensing, while stronger alignment drives large-scale shape changes.

We next investigated how nematic ordering and volume constraints govern phase behavior. The interplay between nematic interaction strength and reduced volume reveals distinct phases (Figures \ref{fig:fig9}, \ref{fig:fig12}). A membrane half-covered with arc-shaped CMCs can exhibit both prolate and oblate phases, depending on the binding strength between CMCs; the prolate and oblate phases occur at low and high nematic interaction strengths, respectively. Tubular structures stabilized by arc-shaped CMCs are well documented in numerous numerical studies \cite{BobrovskaPLOS2013,jarin2020lipid,kralj2000stable,noguchi2025curvature}, and are also supported by experiments with giant unilamellar vesicles \cite{ugarte2019drp1}, curvature-stabilizing proteins in the endoplasmic reticulum (ER) \cite{hu2011weaving}, and in the daughter vesicles of the erythrocyte membrane \cite{bohinc2006shape}. We found that arc-shaped CMCs, in the absence of neighbor interactions, accelerate the oblate-to-prolate transition compared to bare membranes where this transition is driven only by volume constraints (Figure \ref{fig:fig_10}), highlighting how CMC entropy penalizes flat vesicle regions. This entropy-driven effect is consistent with the clustering of CMCs on curved rims (Figure \ref{fig:depletion}), resembling lipid raft aggregation in Golgi cisternae \cite{iglicFEBS04} or pore stabilization \cite{fovsnaric2003stabilization,iglivc2008stabilization}. In contrast, saddle-like CMCs exhibit metastability and neck phases (Figure \ref{fig:fig12}), similar to tubulation phenomena \cite{fovsnaric2005influence}. The persistence of protrusions even without volume constraints (Figure \ref{fig:fig12}(a)) underscores the role of saddle-like anisotropic CMCs in stabilizing connecting necks with negative Gaussian curvature.

At concentrations below 10\% CMC and high interaction strength, we found that saddle-shaped CMCs preferentially accumulate in the elongated necks separating convex empty membrane regions. This observation agrees with previous studies \cite{iglivc2007role,hagerstrand2006curvature} suggesting that vesicle necks contain high concentrations of anisotropic membrane components with non-zero deviatoric curvature. The accumulation of these components in membrane necks correlates with strong in-plane nematic ordering, occurring where the difference between principal membrane curvatures (and consequently, Gaussian curvature) is maximized.

We investigated topological defects (TDs) for both anisotropic CMC types. Our results show that TDs consistently localize to regions of high Gaussian curvature (Figures \ref{fig:fig14}, \ref{fig:fig15}). The +1/2 defects on cylindrical caps and +1 defects at protrusion tips (Figure \ref{fig:fig14}) match theoretical predictions \cite{bowick2004curvature,mesarec2016effective}, with their total topological charge satisfying the Poincaré-Hopf theorem. Along tubular protrusions, the deviatoric curvature term dominates, enforcing CMC alignment parallel to the tube axis, while intrinsic interactions primarily determine defect positioning. This dual behavior suggests that CMCs may utilize both curvature sensing and generation mechanisms in biological systems, as observed in ER tubules \cite{guven2014terasaki} and photoreceptor discs \cite{chandler2008intrinsic}.

Our use of area and volume constraints in combination with a nematic in-plane field is not the first \cite{kumar2019tubulation, behera2021deformation}. However, the systematic exploration of the phase diagram is, to the best of our knowledge, the first to be published in the phase space of binding strength and reduced volume. Also novel is our investigation of the local nematic order $S_i$. The role of TDs was explored by Ramakrishnan and colleagues \cite{ramakrishnan2012role}, but our investigation of the local nematic order $S_i$, particularly during phase transitions, is to the best of our knowledge, the first of its kind. We found that the average mean curvature $H_m$ of the CMCs controls the radius of the pearls and that nematic order in the thin necks is zero. When the CMCs are saddle-shaped, however, and strongly interact, the necks between the pearls are wider and determined by the mean deviator $D_m$ — even when the coverage of CMCs is lower than 10\%. In the latter case, the nematic order in the necks is extremely high. Additionally, such neck-like phases are observed also when the deviatoric term is increased in the absence of direct interactions. This leads us to conclude that the pearling phase is a ubiquitous process which always requires some CMCs to develop, but the ramifications of our simulations could provide insight into the structure of pearling or neck-like phases simply by observing the width of the neck and/or the pearls themselves.

Our findings are in line with the conclusions of \cite{vyas2022sorting}, who emphasized the importance of incorporating both shape and curvature anisotropy, as well as interaction potentials, in understanding protein sorting behavior. Our study demonstrates that curvature anisotropy and interaction strength enhance sorting efficiency, while shape anisotropy can counteract it. In our model, the in-plane nematic field was also researched in the context of non-convex CMCs. Moreover, in-plane coupling to membrane curvature can drive the emergence of complex structures such as tubes and discs, reminiscent of those induced by curvature-sensing proteins \cite{ramakrishnan2012role}. Additionally, our findings support and extend prior work showing that anisotropic curvature-inducing proteins, modeled as in-plane nematics, can drive membrane remodeling and aggregation via membrane-mediated interactions alone, without explicit self-interactions \cite{ramakrishnan2013membrane}. Specifically, the variation of the Gaussian modulus has previously been reported to affect shape changes even in the absence of direct CMC interactions, and its increase can facilitate neck formation even in axis-symmetrical 2D systems \cite{walani2014anisotropic}. Our article highlights the necessity of incorporating anisotropic spontaneous curvature into membrane models to accurately capture vesiculation phenomena, as also demonstrated elsewhere \cite{xiao2023vesiculation}.  

Our simulations establish a framework for understanding how anisotropic CMCs (including BAR proteins and lipid rafts) shape cellular membranes. The boomerang phase (Figure \ref{fig:fig9}) and neck-driven elongation (Figure \ref{fig:fig13}) potentially model \textit{Vibrio cholerae} cells or budding of yeast cells, respectively \cite{alberts2015essential}. However, the metastable behavior of saddle-like CMC phases (Figure \ref{fig:fig12}) requires further investigation, particularly to distinguish between invaginated and elongated phases.

\section{Conclusion}

In this work, we systematically investigated how anisotropic curved membrane components (CMCs) govern vesicle morphology through curvature sensing and nematic alignment. By focusing on arc- and saddle-shaped CMCs, we uncovered a far richer spectrum of membrane shapes than previously observed with isotropic CMCs, where only budding was reported. Our simulations revealed that nematic ordering plays a crucial role in shaping vesicles: for arc-shaped CMCs, weak alignment allows pearling, while stronger alignment stabilizes smooth cylindrical phases. These findings confirm long-standing theoretical predictions and align with experimental studies of BAR domain proteins and tubule-forming systems.

We mapped the morphological transitions as a function of nematic interaction strength and reduced volume, demonstrating how these two parameters control the emergence of prolate, oblate, tubular, and metastable necked morphologies. Notably, arc-shaped CMCs can drive shape changes even in the absence of direct interactions, with entropy playing a key role in destabilizing flat membrane regions and facilitating neck formation. Saddle-shaped CMCs, on the other hand, induce and stabilize negatively curved necks, even at low concentrations and without volume constraints, reinforcing their role in neck formation observed in biological systems.

Our investigation of topological defects (TDs) and local nematic order provides a novel perspective on the coupling between curvature, nematic alignment, and vesicle topology. We also compared our curvature-energy framework with prior models, emphasizing our focus on mean and deviatoric curvature, which better aligns with biological observations of anisotropic curvature sensing. Our phase diagrams are, to our knowledge, the first to fully chart this behavior in the space of binding strength and reduced volume.

Altogether, our findings underscore the importance of anisotropic spontaneous curvature in modeling membrane remodeling and suggest a unifying framework for understanding how proteins such as those containing BAR domains, lipid rafts, and neck-stabilizing proteins shape cell membrane vesicle topology \textit{in vivo}.



\section{Acknowledgements}
Slovenian Research and Innovation Agency core founding No. P2-0232, P3-0388,  projects No. J2-4427, J2-4447, and J3-60063, the University of Ljubljana interdisciplinary preparative project Nanostructurome 802-12/2024-5, and the European Union’s Horizon 2020 Research and Innovation Programme under the Marie Sk\l{}odowska–Curie Staff Exchange project “FarmEVs” (grant agreement no: 101131175). N.S.G. is supported by the Lee and William Abramowitz Professorial Chair of Biophysics (Weizmann Institute) with additional support from a Royal Society Wolfson Visiting Fellowship (U.K.), and acknowledges support by the Israel Science Foundation (Grant No. 207/22). The views and opinions expressed in this publication are solely those of the authors and do not necessarily reflect those of the European Union. Neither the European Union nor the granting authority can be held responsible. 

\section{Appendix}

\subsection{(a) Monte-Carlo procedure}\label{app:procedure}

The membrane is represented by a set of $N$ vertices that are linked by variable length tethers $l$ to form a closed, dynamically triangulated, self-avoiding two-dimensional network of approximately 2N triangles and with the topology of a sphere \cite{gompper1996,gompper2004}. The lengths of the tethers can vary between a minimal and a maximal value, $l_{min}$, and $l_{max}$, respectively. Self-avoidance of the network is ensured by choosing the appropriate values for $l_{max}$ and the maximal displacement of the vertex $s$ in a single updating step.

One Monte-Carlo sweep (MCs) consists of individual attempts to displace each of the $N$ vertices by a random increment in the sphere with radius $s$, centered at the vertex, followed by $R_B N$ attempts to flip a randomly chosen bond. We denote $R_B$ as the bond-flip ratio, which defines how many attempts to flip a bond are made per one attempt to move a vertex in one MCs. Note that the bond-flip ratio is connected to the lateral diffusion coefficient within the membrane, i.e. to the membrane viscosity. In this work we have chosen $R_B = 3$,  $s/l_{min}=0.15$ and $l_{max}/l_{min}=1.7$. The dynamically triangulated network acquires its lateral fluidity from a bond flip mechanism. A single bond-flip involves the four vertices of two neighboring triangles. The tether connecting the two vertices in diagonal direction is cut and reestablished between the other two, previously unconnected, vertices. The self-avoidance of the network is implemented by ensuring that no vertex can penetrate through the triangular network and that no bond can cut through another bond \cite{samo2015,fovsnarivc2019}.

\subsection{(b) Anisotropic code details}
\subsubsection{Representing the membrane as a mesh}\label{app:code details}

Our solver is called \texttt{Trisurf}. It models the vesicle as a closed, triangulated surface: a graph with vertices $i\in V$ and edges $e_{ij}\in E$, and an auxiliary set of triangles $t_{ijk}\in T$. The triangles make the approximation of the surface, but it is the vertices which are the principal dynamical entities that hold the properties of the membrane (intrinsic curvature, membrane composition, nematic director, etc.) and move in space. 

From the position of the vertices, $\vec{x}_i $ we can compute the normal vector to each triangle $\vec{N}_{ijk}$ and circumcenter $\vec{O}_{ijk}$ (center of the circle containing the position of $\vec{x}_i,\vec{x}_j,\vec{x}_k$). This allows us to divide the triangle to six parts and assign two pieces to each vertex.

\begin{figure}[h]
    \centering
    \includegraphics[width=0.4\linewidth]{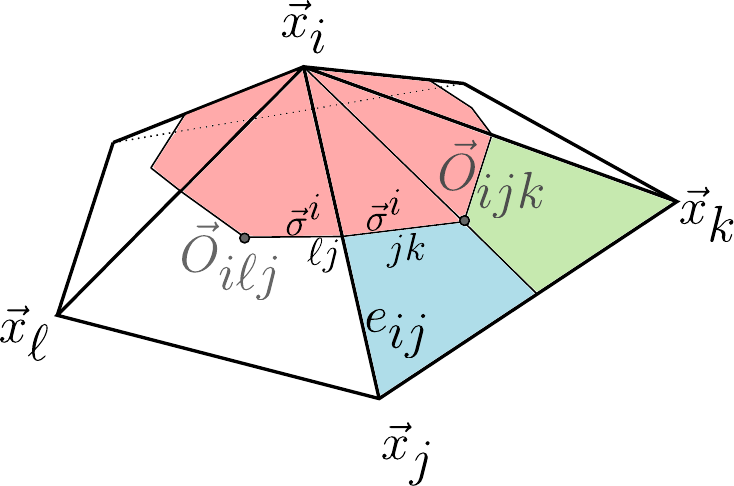}
    \caption{Schematic of a vertex and it's neighbors. The vertex $i$ has neighbors, $\ell,j,k,\ldots$, which are connected by edges $e_{ij},\ldots$, where we explicitly show $e_{ij}$. Each triangle has a circumcenter $O$, and the area is assigned to each of the 3 vertices, which is shown for triangle $ijk$ (red, green, and blue section). The total area assigned to $i$ is shown in red across the entire cap, and the dual edge $\sigma^{i}$ accros $e_{ij}$ are shown $\sigma^{i}_{jk}$ and $\sigma^i_{\ell j}$. The curvature of vertex $i$ is essentially computed on the red area, for example, the normal of the vertex $N\left(i\right)$ is the average of the local normal for every point in the red surface.}
    \label{fig:Area_assignment}
\end{figure}

For vertex $i$, these are the sub-triangle between the vertex position $\vec{x}_i$, the triangle center $\vec{O}_{ijk}$, and the middle of one of the edges $\frac{\vec{x}_i+\vec{x}_j}{2}$, and a similar sub-triangle for the other edge $ik$. We can denote the vector for the side between the vertex and the edge middle as half the edge length $\vec{e}_{ij} =\vec{x}_j-\vec{x}_i$ and the side between center and the edge middle, which is half of the dual (voronoi) edge, as $\vec{\sigma}^{i}_{jk}$ (the other half of the voronoi edge is on the neighboring triangle $\sigma^{i}_{\ell j}$). We can compute this $\sigma$ from the circumcenter and the middle of $\vec{x}_i,\vec{x}_j$. This allows us to divide the triangle to six parts and assign two pieces to each vertex (Figure \ref{fig:Area_assignment}).

\begin{equation}\label{eq:VoronoiEdge}
    \vec{\sigma}^{i}_{jk} = \frac{\vec{x}_i+\vec{x}_j}{2} - \vec{O}_{ijk} 
\end{equation}

With this, we assign an area $A\left(i\right)$ and a normal $N\left(i\right)$ for each vertex $i$, by running over the neighbors $j-1,j,j+1\ldots$ and the adjacent triangles $\left(i,j-1,j\right),\left(i,j,j+1\right)\ldots$
\begin{equation} \label{eq:VertexArea}
    A\left(i\right) = \sum_{\left\langle i,j\right\rangle} \frac{1}{2} \frac{\left|e_{ij}\right|}{2}\left|\sigma^i_{j,j+1}\right| + \frac{1}{2} \frac{\left|e_{ij}\right|}{2}\left|\sigma^i_{j,j-1}\right|
\end{equation}
\begin{equation} \label{eq:VertexNormal}
    \vec{N}\left(i\right) = \frac{\sum_{\left\langle i,j\right\rangle}  \vec{N}_{ij,j+1} \left|e_{ij}\right|\left|\sigma^i_{j,j+1}\right| +  \vec{N}_{i,j-1,j} \left|e_{ij}\right|\left|\sigma^i_{j,j-1}\right|}{\left|\ldots\right|}
\end{equation}
Where the $j$ vertices are the counterclockwise ordered neighbors of $i$.

The fluidity of the surface is achieved by bond-flips, where a bond $ij$ and the two triangles that share it is $ijk,ji\ell$ are replaced by a cross bond $k\ell$ and two triangles $i\ell k,j k \ell$, which allows vertices to change neighbors.

\subsubsection{Anisotropic curvature on the vesicle}

we use the shape operator, which is a discreet version of the principal curvatureas on a mesh, but we calculate it by projecting it  onto a plane d (director) and t (tangent) to get the 2x2 matrix C(v). By this, we have 2H given by tr(C) and det(C) gives K

To get the anisotropic bending energy of the surface, we use the method by \cite{ramakrishnan2010monte} to estimate the shape operator matrix $S$ on each vertex, which represents the shape of the surface at the point. We then calculate the mismatch tensor $M=S-C_m$, where $C_m$ is the intrinsic curvature tensor whose direction is determined by the director and the other tangent vector to the local normal ($\hat{t}=\hat{N}\times\hat{d}$). 
\begin{equation}
    C_m = \frac{H_m+D_m}{2}\hat{d}\otimes\hat{d} + \frac{H_m-D_m}{2} \hat{t}\otimes\hat{t}
\end{equation}
where $H_m=(C_{1m}+C_{2m})/2$, $D_m=(C_{1m}-C_{2m})/2$ are the spontaneous curvature and spontaneous deviator at the vertex, respectively, which reflects the physical characteristics of local membrane composition. This is a change from the original method by Ramakrishnan et al. \cite{ramakrishnan2014mesoscale}, where the $P_\textbf{v}$ was used as a matrix projection, and could be the detail that makes the calculations more robust and efficient.

The bending energy is calculated by inserting the mismatch tensor in the Hamiltonian
\begin{equation}\label{eq:8}
    E_1=\frac{K_1}{2}\left(\mathrm{Tr} M\right)^2 + K_2 \mathrm{Det}M
\end{equation}
Where $K_1$ and $K_2$ are the bending moduli of the vertex, again reflecting physical parameters due to local composition.

To calculate the shape curvature of a vertex $i$ based on \cite{ramakrishnan2010monte}, each edge $ij$ is assigned a shape tensor estimation
\begin{equation}
    S_{ij}=h_{ij} \; \vec{b}\otimes \vec{b}
\end{equation}
$\vec{b}$ is the binormal at the edge $\hat{N}_{ij}\times \vec{e}_{ij}$, where $\vec{e}_{ij} = \vec{x}_j-\vec{x}_i$ is the edge vector and $\hat{N}_{ij}$ is the normal of the edge $\hat{N}_{ij} = \left(\hat{N}_{i,j-1,j} + \hat{N}_{i,j,j+1} \right)/ \left|\ldots\right| $ which is the sum of the normal of the two triangles sharing the edge, normalized.
$h_{ij}$ is a factor representing the directional derivative of the area $\approx \nabla_p A$
\begin{equation}
    h_{ij} = 2 \left|e_{ij}\right|\cos\left(\frac{\Phi}{2}\right)
\end{equation}
Where $\Phi$ is the dihedral angle (angle between the two triangle sharing the edge).
Luckily there is a simple triple product formula for this factor
\begin{equation}
    h_{ij} = 2\left|\vec{e}_{ij}\right| \vec{N}_{ij} \cdot \left(\vec{N}_{i,j-1,j} \times \hat{e}_{ij} \right)
\end{equation}
The cross product of the edge direction and a triangle normal gives a vector on the triangle which is perpendicular to the edge, which is at an angle $\frac{\Phi}{2}$ from the edge normal. 

The full vertex-shape tensor is a sum of the edge tensor, weighted by the match of the normal of the edge to the normal of the vertex.
\begin{equation}
    S\left(i\right) = \frac{1}{A\left(i\right)}\sum_j \hat{N}\left(i\right)\cdot\hat{N}_{ij} \; S_{ij}
\end{equation}

We then project this 3x3 matrix in real space $\hat{x},\hat{y},\hat{z}$ to the tangent plane of the vertex $\hat{d},\hat{t}$
\begin{equation}
    S_{2\times2} =\begin{pmatrix}
        \hat{d}\cdot S \cdot \hat{d} & \hat{d} \cdot S \cdot \hat{t} \\
        \hat{t}\cdot S \cdot \hat{d} & \hat{t} \cdot S \cdot \hat{t}
    \end{pmatrix}
\end{equation}

The mean curvature $H$ at the vertex is half the trace, while the Gaussian curvature $K$ is the determinant, which are two of the degrees of freedom in the Hamiltonian.

The mismatch matrix can be calculated
\begin{equation}
    M = \begin{pmatrix}
        S_{dd} - \frac{H_m + D_m}{2}& S_{dt} \\
        S_{td} & S_{tt} -  \frac{H_m - D_m}{2}
    \end{pmatrix}
\end{equation}
The angle between the director and the eigenvectors of the shape matrix is what results in the $\omega$ angle dependence, which is the final degree of freedom.
The mismatch tensor is simply inserted into Eq. (\ref{eq:8}) to compute the bending energy of the vertex. If we integrate it across the whole membrane, we get Eq. (\ref{eq:eq1}).

Let us shortly comment the main differences in numerical methodology. While Kumar and colleagues \cite{kumar2022review} project the principal curvatures — and their corresponding bending rigidities — along both principal directions in the tangent plane of each membrane vertex, we use their sum and difference, namely the mean curvature and mean deviator (and correspondingly, bending and splay stiffness). When comparing the two approaches, one might notice a discrepancy in the absolute value of the energies, but this difference is not important for the method itself, as the Monte Carlo steps and the convergence of both methods rely only on the energy difference between the former and latter states, $\Delta E$. Our methodology is distinct because we explicitly apply the theoretical results posited nearly thirty years ago in the works of Kralj-Iglič and Iglič \cite{iglivc2007,iglivc2005role,kralj2002deviatoric,kralj2012stability}. To make this example more concrete, let us directly compare our method with that of Kumar and colleagues \cite{kumar2019tubulation} for a simple case of a tubulating membrane. To investigate the steady-state shapes of membranes fully covered by nematic CMCs, the method of \cite{kumar2019tubulation} uses $c_{||}$ and $c_{\perp}$, while our method uses $H_m$ and $D_m$. For an outward tubulation of the membrane, the first method uses $c_{||}<0$ and $c_{\perp}>0$, while our method uses $H_{m}=D_{m}>0$. Similarly, for inward tubulation, the first method uses $c_{||}>0$ and $c_{\perp}<0$, and our method $H_{m}=D_{m}<0$. Of course, the length scales of the CMCs relative to the membrane size would need to be adjusted and matched in both cases. Anisotropic CMCs interact with both principal curvatures of the membrane, so in any case, the degrees of freedom are two; what differs is the parametrization.

\subsection{(c) Supplementary Information}
\subsubsection{Pearling transition up close}

\begin{figure}[h]
    \centering
    \includegraphics[width=\linewidth]{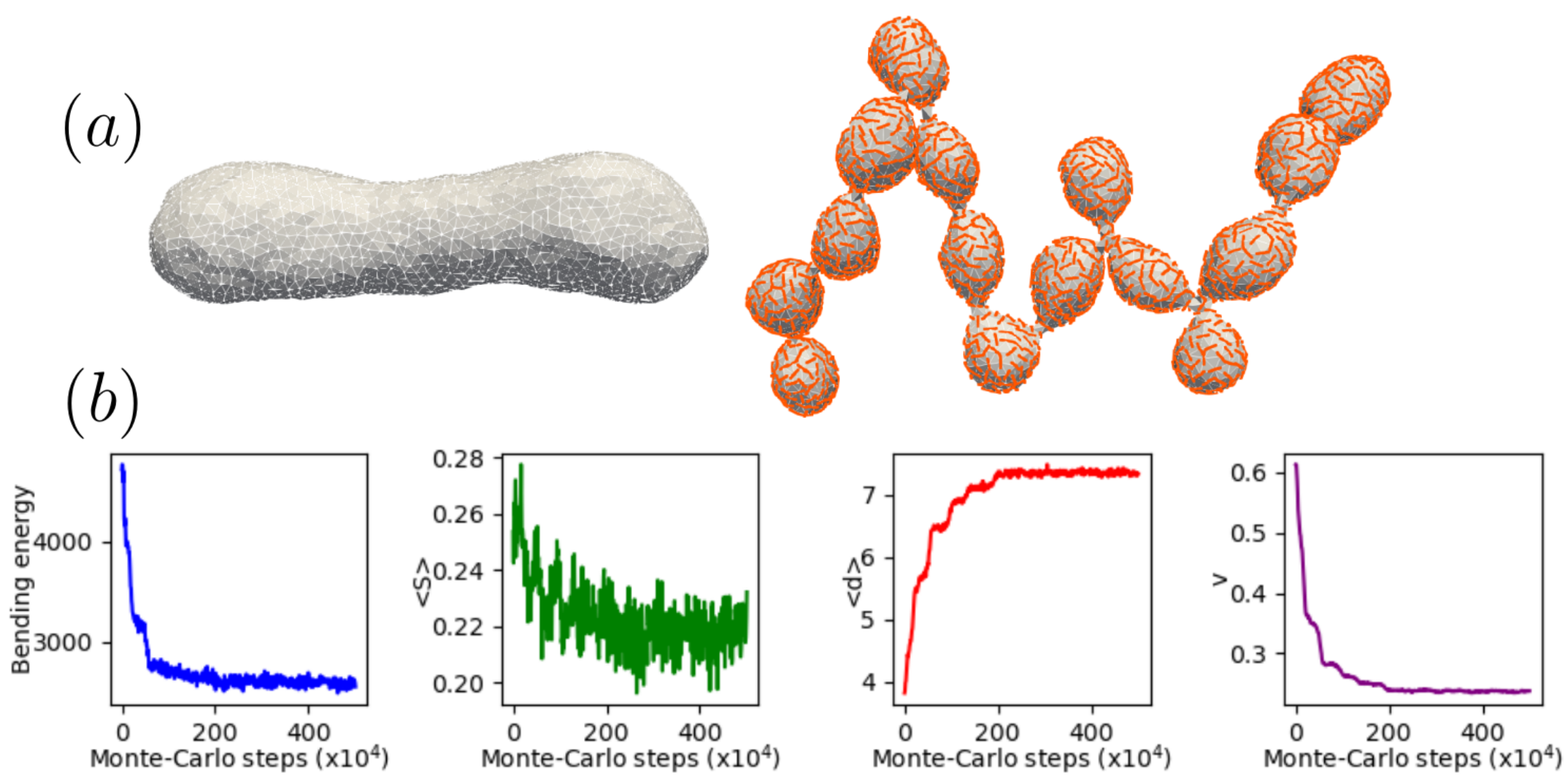}
    \caption{(a) Populating an empty cylinder with arc-shaped CMCs ($H_m=D_m=0.5$) and setting $w=0$ results in pearl steady-states shapes. (b) This transition is accompanied by a drop in bending energy, average nematic order and volume, but an increase in average deviator.}
    \label{fig:supp1}
\end{figure}

We study the transition from cylinders to pearl-like steady-state shapes by populating a previously empty cylinder with arc-shaped CMCs and running the simulation (Figure \ref{fig:supp1}).

\subsubsection{Pearling transition by varying $H_m$}

At $w=0$ and changing $H_m$, the steady-state shape changes from a sphere to a pearling state as shown in Figure \ref{fig:supp2}.

\subsubsection{Spontaneous ordering in necks when $K_2$ increases}

For saddle-like CMCs at $w=0$ and increasing $K_2$ (see Eq. (\ref{eq:eq1})), the steady-state shapes form saddle-like necks between empty convex membrane regions (see Figure \ref{fig:fig13}). This is due to the dominance of the deviatoric term.

\begin{figure}[h]
    \centering
    \includegraphics[width=0.5\linewidth]{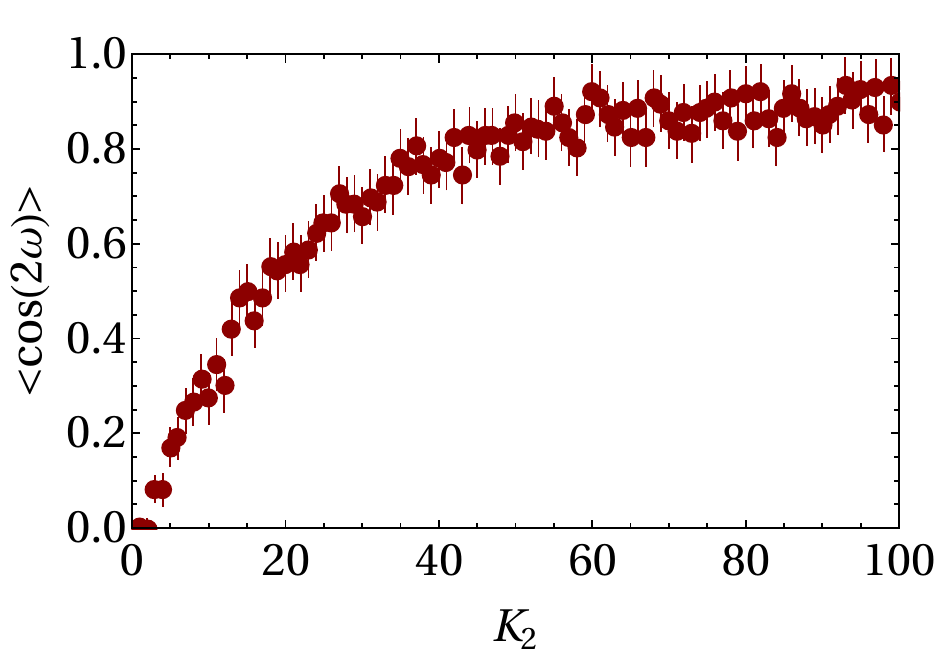}
    \caption{(a) With no nematic interaction ($w=0$), but increasing $K_2$, the steady-states of vesicles form saddle-like necks between alternate convex regions and show perfect alignment between neighboring saddle-like CMCs at low concentrations. The $x$-axis is the constant $K_2$ in Eq. \ref{eq:eq1}, while $\langle\cos{(2\omega)}\rangle$ is the average orientation between neighboring CMCs. Parameters here are for simulations with $H_m=0$, $D_m=0.98$, $\rho=0.06$, $w=0$.}
    \label{fig:supp2}
\end{figure}

\begin{figure}[h]
    \centering
    \includegraphics[width=1.1\linewidth]{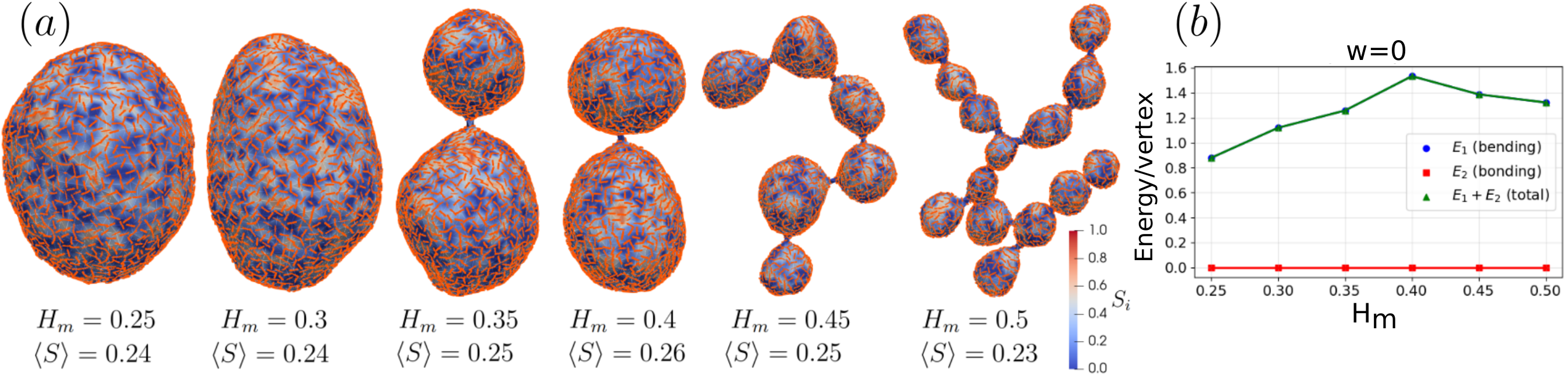}
    \caption{(a) With no nematic interaction ($w=0$), but varying $H_{m}$, the steady-states of vesicles gradually change to pearls. (b) Total energy as a function of $H_m$ shows that bending energy initially increaes, but is reduced once the number of pearls is exceeds 2.}
    \label{fig:supp2}
\end{figure}

\subsubsection{Arc-shaped CMCs cluster on the rim to form oblates even in absence of nematic ordering}\label{app:supp2}

\begin{figure}[h!]
    \centering
    \includegraphics[width=0.8\textwidth]{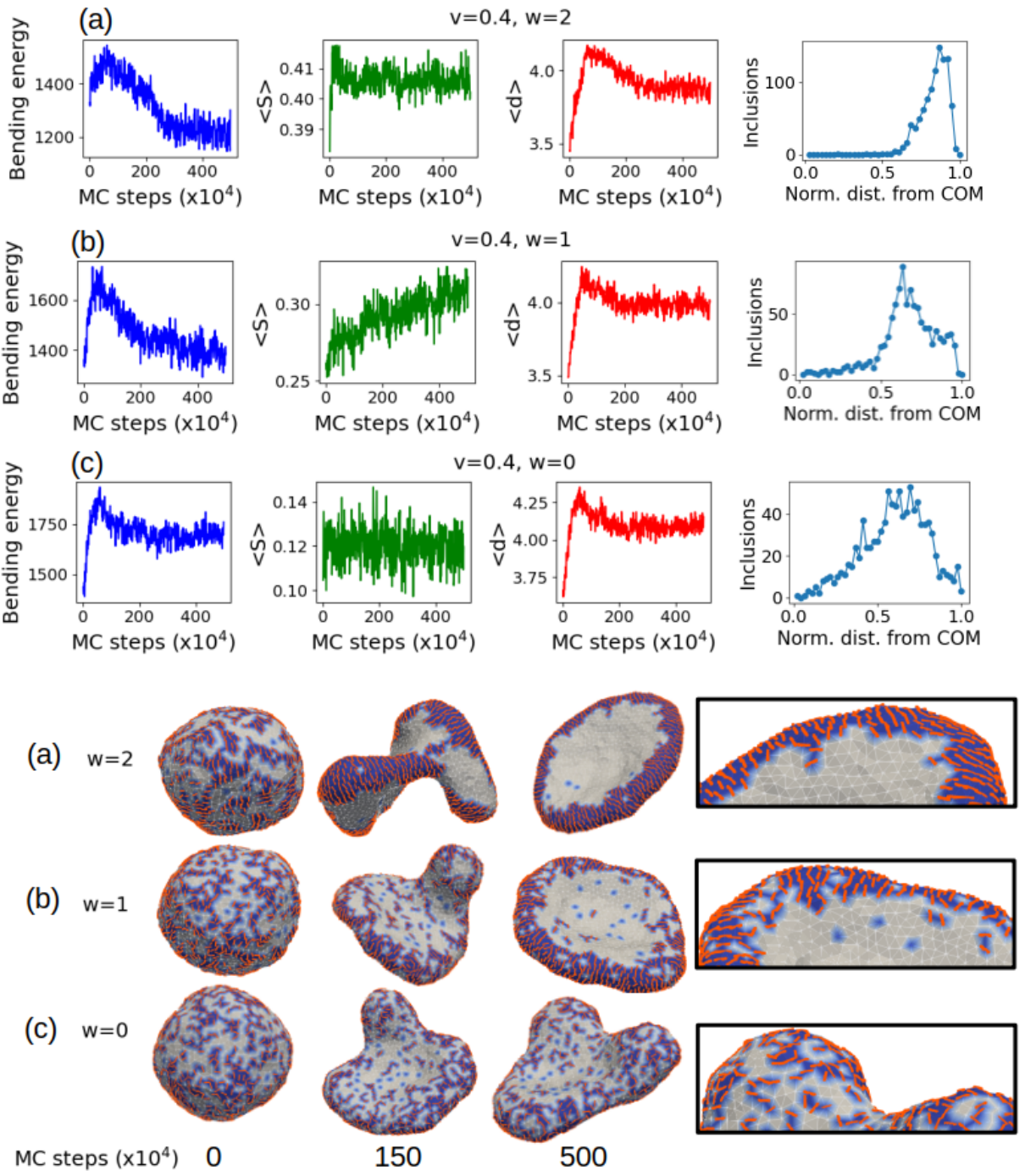}
    \caption{Oblate shapes $v=0.4$ at $\rho=0.5$ of arc-shaped CMCs ($H_m=D_m=0.25$) and three values of $w$ (from 0 to 2). Even with no nematic interaction between CMCs (c), these are not distributed homogenousely over the membrane, but tend to accumulate on the rim, as is reflected in the inclusion distribution from the COM (row (c)). This slight ordering results in a positive average nematic order $\langle S \rangle$. Slight nematic ordering of $w=1$ (b) results in the majority of CMCs being located at the rim (row (b)) and nematic order $\langle S \rangle$ increasing in the process. Deviator and radial dependence for $H_m=D_m=0.25$. The CMCs profile from the centre of mass (COM) is made for the last state, namely at 500 MC steps.}
    \label{fig:deviator}
\end{figure}

Figure \ref{fig:deviator} shows the total energy, average nematic order, average deviator and CMC distribution during the sphere-oblate transition for $\rho=0.5$ of arc-shaped CMCs ($H_m=D_m=0.25$) at $v=0.4$ and three values of $w$. Vesicle energy (equation \ref{eq:eq2}) converges to a minimum in all three cases, but is lower at greater values of $w$. Average nematic order is largest for $w=2$ and lowest for $w=0$ when no explicit nematic ordering is present. 

The average curvature deviator for each shape is calculated and shown in the third column of Figure \ref{fig:deviator}. The membrane deviator is defined for each vertex as $D=(C_1 - C_2)/2$, where $C_1$ and $C_2$ are its two principal curvatures. Dimensionless deviator is given by $d=RD$, where $R$ is the radius of the sphere with the same area as the vesicle. The average deviator for each shape is calculated as $\langle d \rangle = \int d \, da/\int da$ \cite{kralj2020minimizing}. We find that the deviator slightly decreases with increasing $w$, as shown in the third column of Figure \ref{fig:deviator}; all shapes are oblates with varying degrees of flatness, with most flat regions corresponding to $w=2$, where -- due to strong nematic ordering -- most of the CMCs are located at the rim (see the upper figure of the fourth column of Figure \ref{fig:deviator}). Note that a sphere would have $d=0$.   

\printbibliography  

\end{document}